\documentclass[british,prstab,groupedaddress,amsmath,reprint,floatfix]{article}
\usepackage[T1]{fontenc}
\usepackage[latin9]{inputenc}
\usepackage{geometry}
\usepackage{xcolor}
\usepackage{pdfcolmk}
\usepackage{babel}
\usepackage{array}
\usepackage{float}
\usepackage{units}
\usepackage{textcomp}
\usepackage{mathrsfs}
\usepackage{enumitem}
\usepackage{amsmath}
\usepackage{amssymb}
\usepackage{graphicx}
\usepackage{esint}
\PassOptionsToPackage{normalem}{ulem}
\usepackage{ulem}
\usepackage[unicode=true]
 {hyperref}

\makeatletter

\newcommand{\lyxmathsym}[1]{\ifmmode\begingroup\def\b@ld{bold}
  \text{\ifx\math@version\b@ld\bfseries\fi#1}\endgroup\else#1\fi}

\providecommand{\tabularnewline}{\\}
\providecolor{lyxadded}{rgb}{0,0,1}
\providecolor{lyxdeleted}{rgb}{1,0,0}
\DeclareRobustCommand{\lyxsout}[1]{\ifx\\#1\else\sout{#1}\fi}

\newlength{\lyxlabelwidth}      % auxiliary length 
\newcommand{\lyxaddress}[1]{
	\par {\raggedright #1
	\vspace{1.4em}
	\noindent\par}
}
	\newenvironment{elabeling}[2][]%
	{\settowidth{\lyxlabelwidth}{#2}
		\begin{description}[font=\normalfont,style=sameline,
			leftmargin=\lyxlabelwidth,#1]}
	{\end{description}}

\usepackage{cite}
\usepackage{tocloft} 
\usepackage{siunitx}

\DeclareMathOperator{\atan2}{atan2}

\newcommand{\totd}{\,\mathrm d}

\usepackage{mathrsfs}
\usepackage{revsymb}

\makeatother

\begin{document}
\title{A CAD Tool for Linear Optics Design:\\
A Controls Engineer's Geometric Approach to Hill's Equation}
\author{J. Bengtsson\textsuperscript{1}, W. Rogers\textsuperscript{2}, T.
Nicholls\textsuperscript{2}}
\maketitle

\lyxaddress{\begin{center}
\textsuperscript{1}Helmholtz-Zentrum Berlin, BESSY, Berlin, Germany\\
\textsuperscript{2}Diamond Light Source, Ltd, Oxfordshire, UK
\par\end{center}}
\begin{abstract}
\textit{The formulae relevant for the Linear Optics design of Synchrotrons
are derived systematically from first principles, a straightforward
exercise in Hamiltonian Dynamics. Equipped with these, the relevant
``Use Cases'' are then captured for a streamlined approach. This
will enable professionals - Software Engineers \textendash{} to efficiently
prototype \& architect a CAD Tool for ditto; something which has been
available to Mechanical Engineers since the mid-1960s.}

\noindent \textit{\newpage}
\end{abstract}
\tableofcontents{}\newpage{}

\section{Introduction}

This paper has two objectives:
\begin{enumerate}
\item To capture \& outline the Linear Optics Design ``Use Cases''. For
a straightforward \& streamlined systematic approach useful for ``rapid
prototyping'' \& implementation, e.g. \cite{Use_Case_1,FLAME,Use_Case_2,Use_Case_3},
of in this case a CAD Tool for Linear Optics Design. Something that
has been available to e.g. Mechanical Engineers since the mid-1960s.
\item To enable the pursuit of (1): to outline \& summarize the Mathematical
Framework for Linear Optics Design for a charged relativistic particle
beam in an external electromagnetic field, in a concise yet comprehensive
fashion; i.e., from first principles \& self-consistently.\\
This is an exercise in Hamiltonian Dynamics; perturbed by Classical
Radiation \& Quantum Fluctuations.\\
Hamiltonian Dynamics was developed almost two centuries ago \cite{Lagrange_0,Lagrange,Hamilton_1,Hamilton_2};
a century before the invention of the digital computer \cite{Turing,von_Neumann}.\\
More generally, as a first step to re-introduce an analytic approach
for Robust Lattice Design; vs. the contemporary brute force, ``black
box'', and ad hoc numerical pursuits.\\
In particular, for Chasman-Green type lattices \cite{Chasman_Green},
for which the ``Theoretical Minimum Emittance'' Cell (TME), a reductionist's
concept \& view, which missed the virtue of Reverse Bends \cite{Streun}
when introduced to Damping Rings in the late 1980s \cite{Reverse_Bend}.
For a systematic approach for the latter, see ref. \cite{Raubenhemier}.\\
For a brief summary of an in-depth analytic approach to \textendash{}
the more than a century old ``Main Problem'' \textendash{} the 3-Body
Problem: Earth, Moon, and Sun \cite{Delaunay,Deprit_1,Deprit_2,Hill},
see intro p. 34, section 7 in ref. \cite{CERN-88-05}.
\end{enumerate}
It is known how to develop software tools using Best Practices (reusable,
modular, extensible, etc.); see e.g. ref. \cite{GNU}. In particular,
to enable\footnote{Tracy-2->3->4 was implemented as a Modular, Software Library, by the
first author in Pascal in the early 1990s; for a state-of-the-arts
beam dynamics model at the time. Initially, to guide the ALS commissioning
\cite{Tracy-2_ALS}. It was used for the conceptual design \& commissioning
of e.g. the SLAC B-Factory\cite{PEP-II}, SLS \cite{Tracy_2_SLS_1},
SOLEIL \cite{Tracy-2_SOLEIL}, and DIAMOND \cite{Tracy-2_DIAMOND_1,Tracy-2_DIAMOND_2}
as well.}:
\begin{itemize}
\item reusing the Beam Dynamics Model \& related Controls Algorithms guiding
the Conceptual Design as a Model Server for Model-Based Control of
the Commissioning \cite{Tracy-2_SLS_2} (first by automated machine
translation from Pascal to C (with p2c \cite{p2c}), and then implemented
as a Shared Resource by introducing a Client/Server Architecture),
\item implementing a ``Virtual Accelerator'' for End-to-End Testing of
Controls Applications; before Commissioning the real system \& facility
\cite{Tracy-2_DIAMOND_2,Tracy-2_NSLS-II},
\item implementing a MATLAB Interface (aka Accelerator Toolbox) \cite{Tracy-2_AT},
\item implementing a Python Interface \cite{Tracy-2_Python,Tracy-2_Cython}.
\end{itemize}
A mathematical model is provided by the Lorentz Force, i.e., the Equations
of Motion are
\begin{equation}
\frac{d\bar{p}}{dt}=q\left(\bar{E}+\bar{v}\times\bar{B}\right),\qquad\bar{p}\equiv m_{0}\gamma\bar{v},\qquad\gamma\equiv\frac{1}{\sqrt{1-\nicefrac{v^{2}}{c_{0}^{2}}}}
\end{equation}
where $\bar{p}$ is the relativistic momentum, $m_{0}$ the rest mass,
$\gamma$ the relativistic factor, $c_{0}$ the speed of light in
vacuum, $\bar{v}$ the velocity, $q$ the charge, and $\bar{E},\bar{B}$
the electric \& magnetic fields, respectively. In the following considerations
$\bar{E}=0$, i.e., the beam is guided by magnetic fields, apart from
the RF Cavity, for which $\bar{E}\neq0,\bar{B}=0$; to compensate
for the particles energy loss by radiation on a turn-by-turn basis.
For ultra-relativistic beams, $\bar{v}\rightarrow c_{0}$, the particles'
energy \& momentum are essentially the same 
\begin{equation}
E=\gamma m_{0}c_{0}^{2}=\sqrt{\left(m_{0}c_{0}^{2}\right)^{2}+\left(pc_{0}\right)^{2}}\approx pc_{0},\qquad\gamma\gg1.\label{eq:rel_energy}
\end{equation}
For DIAMOND, storing electrons at $E=3$ GeV with a rest mass of $m_{0}c_{0}^{2}=0.511$
MeV with $\gamma=\nicefrac{E}{m_{0}c_{0}^{2}}=5.0\times10^{3}$; the
difference is only $c_{0}-v=4.4$ m/s.\\
N.B.: After the magnetic guiding field has been determined, it is
straightforward to numerically integrate the Equations of Motion,
aka Ray Tracing or Tracking, with e.g. a 4th order Runge-Kutta \cite{Runge,Kutta}.
However, long term tracking, for e.g. one damping time or more, requires
care; i.e., using a Symplectic Integrator \cite{Yoshida}, which preserves
the (differential) geometric properties of the phase space flow.\newpage{}

The paper is organized as follows:
\begin{itemize}
\item Section \ref{sec:Linear-Optics} summarizes the results of objective
(2); i.e., Eqs. \eqref{eq:brho}-\eqref{eq:tune}. Additionally, for
a self-contained approach, the details are provided in Appendices
\ref{sec:Rel_H}-\ref{sec:Rad_Effects}, Eqs. \eqref{eq:Rel_H}-\eqref{eq:ID_impact}.
\item Section \ref{sec:Use-Case_Approach} summarizes the primary Use Cases
for state-of-the-art Linear Optics Design.
\end{itemize}

\section{Linear Optics\label{sec:Linear-Optics}}

For Control Engineering, e.g. ref. \cite{Maxwell}, mathematically/traditionally,
there are two different approaches \cite{Ctrl_Theory_1}:
\begin{enumerate}
\item Classical Control Theory (1950s: Nyquist, Bode, Routh, etc.) \cite{Nyquist,Bode,Routh}:
model the system by a linearized Ordinary Differential Equation (ODE),
introduce a Transfer Function, Fourier expand, introduce a feedback
loop with e.g. a PID Controller, etc.
\item Modern Control Theory (1960s: Kalman, etc.) \cite{Kalman}: introduce
a state space, model the system by a state matrix, and utilize Linear
Algebra to analyze the stability, controllability, introduce a state-space
estimator, and full state feedback.
\end{enumerate}
The first approach is limited by the fact that it is rarely generalized
beyond Single-Input-Single-Output (SISO) Systems to Multiple-Input-Multiple-Output
(MIMO); e.g. orbit feedback systems, see Fig. \eqref{fig:Lin_Ctrl_Theory}.
From this point of view, the classic (in this field) Courant \& Snyder
paper \cite{C-S} \textendash{} for a Controls Engineer \textendash{}
is simply linear stability analysis of the state phase for the system;
i.e., the second approach. For a modern Control Theory approach to
Hill's Equation: state-space estimator, full-state feedback by pole
placement, etc., i.e., what to add/introduce to the R.H.S. of the
Equations of Motion to Control the System: an ``Active'' => ``Proactive''
(vs. ``Passive'' => ''Reactive'') point-of-view \& approach, see
ref. \cite{Ctrl_Theory_2}. An early example of \textit{``Engineering-Science''}
\cite{NASA} is the early days of flight, where the Wright Brothers
were the first to achieve \textit{Sustained Flight}; by a systematic
approach vs. trial-and-error-and-error..., (i.e., a long chain of
mis-guided attempts based on copying birds). This enabled them to:
first discover the Principles of Flight (aka 3-Axis Control), from
wind tunnel testing, and then invent the Ailerons (which was implemented
by ``Wing Warping'' at the time) \cite{Wright_Bros}.

However, the first approach is often used for grinding out ``analytical
formula'', i.e., by working directly on the ODEs, for producing academic
papers; i.e., of limited practical use. Contrarily, practical numerical
\& analytical tools are in general based on the Poincaré Map. The
end result: a dichotomy; produced by ``tunnel vision'' resulting
from reductionism, with the Theoretical Minimum Emittance (TME) Cell
at it's pinnacle:
\begin{itemize}
\item quite successfully ignored for the MAX-IV Conceptual Design, i.e.,
by instead pursuing a Systems Approach \cite{MAX_IV_saga,MAX_IV_DDR,MAX-IV},
\end{itemize}
On the other hand, the Minimum Emittance (ME) Cell:
\begin{itemize}
\item was made practical to guide the design of Damping Rings \cite{Raubenhemier},
\item and, eventually, made obsolete by ``looking outside the box'' \cite{outside_the_box},
i.e., by introducing Reverse Bends for ditto \cite{Reverse_Bend},
\item now utilized for SLS-2 \cite{SLS-2,SLS-2 CDR}.
\end{itemize}
N.B.: The pattern, aka ``paradigm shifts'', was observed by the
philosopher of science T. Kuhn in the 1950s \cite{Kuhn}.

Caveat: because the particles move close to the speed of light, Robust
Design, to e.g. obtain satisfactory Beam Life Time, Injection Efficiency,
etc., for a Storage Ring, is largely based on ``Feedforward''. In
other words, Robust Lattice Design for Predictable Results is essentially
a matter of Systems Approach \& providing (realistic) guidelines for
Engineering Tolerances, etc. In particular, unsurprisingly, the performance
of the System depends on the imperfections vs. academic exercises
(TME Cell, etc.). Exceptions to this include Stochastic \& Electron
Cooling \cite{Stoch_Cooling_1,Stoch_Cooling_2,e-Cooling_1,e-Cooling_2}.
\begin{center}
\begin{figure}[H]
\centering{}\includegraphics[width=6cm]{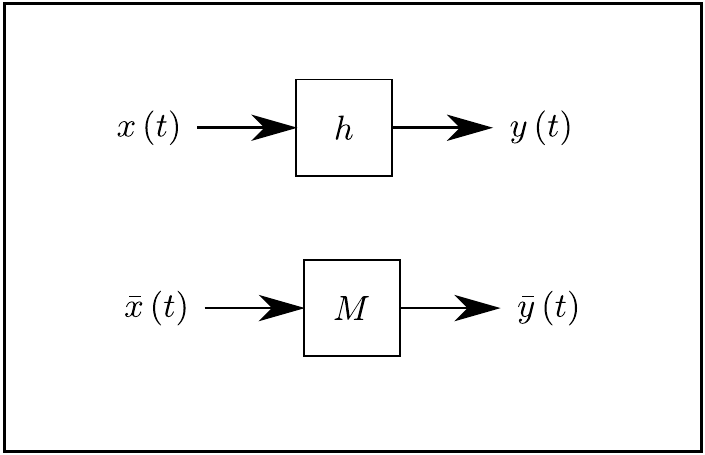}\caption{\textit{Modern} MIMO vs. \textit{Classical} SISO \emph{Control Theory}
\textendash{} \emph{State Space} vs. \textsl{Transfer Function} Approach.\label{fig:Lin_Ctrl_Theory}}
\end{figure}
\par\end{center}

\subsection{Quadratic Hamiltonian \& Linear Maps}

The comoving coordinate system customary for modelling synchrotrons
is shown in Fig. \eqref{fig:Co-Moving_Frame}. In particular, a reference
curve \textendash{} e.g. design orbit for a reference particle \textendash{}
is introduced, and the equations-of-motions for a particles moving
relative to the reference particle are then utilised. Akin to Lagrange
vs. Euler equations for fluid dynamics for which the former describes
how e.g. flotsam in a river moves relative to an observer who is moving
with the stream.
\begin{center}
\begin{figure}[H]
\centering{}\includegraphics[width=6cm]{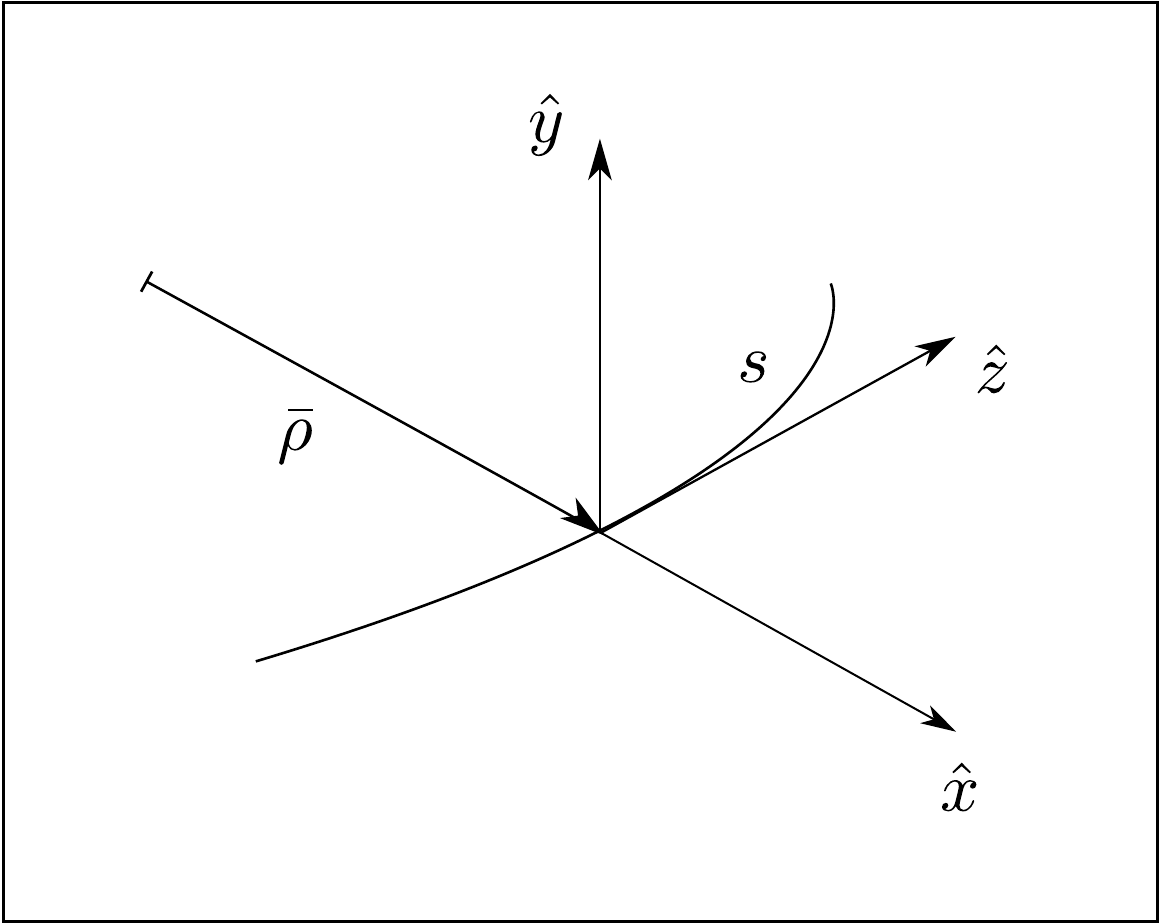}\caption{Co-Moving Frame Coordinates.\label{fig:Co-Moving_Frame}}
\end{figure}
\par\end{center}

The geometry of the trajectory for a charged particle traversing a
magnetic field \textendash{} an arc \textendash{} is summarized by
the Magnetic Rigidity
\begin{equation}
\frac{p}{q}=B\rho,\qquad\phi\equiv\frac{L}{\rho}\label{eq:brho}
\end{equation}
see Fig. \ref{fig:Magn_Rig}.
\begin{center}
\begin{figure}[H]
\begin{centering}
\includegraphics[width=6cm]{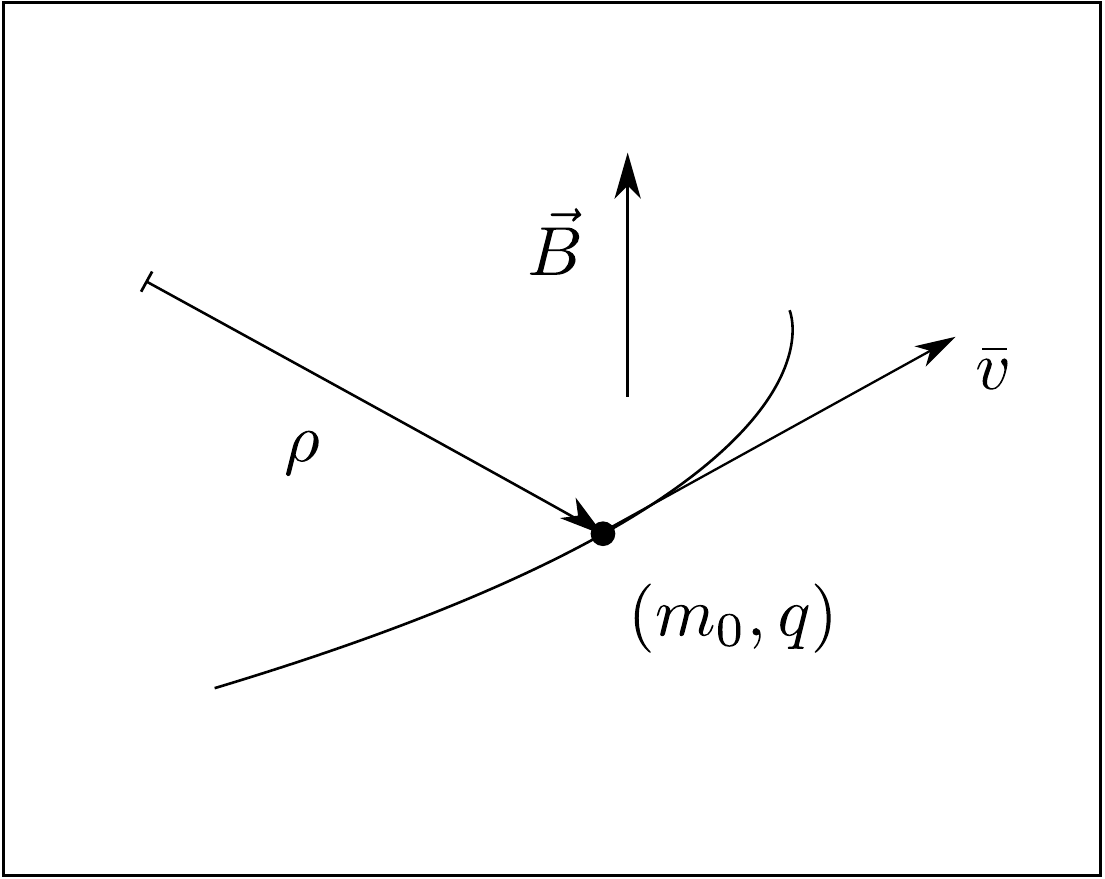}
\par\end{centering}
\caption{Magnetic Rigidity: Geometry\label{fig:Magn_Rig}.}
\end{figure}
\par\end{center}

The magnetic field is described by the Multipole Expansion \cite{SLS_1997}
\begin{align}
B_{y}\left(s\right)+iB_{x}\left(s\right) & =\left(B\rho\right)\sum_{n=1}^{\infty}\left(ia_{n}\left(s\right)+b_{n}\left(s\right)\right)\left(re^{i\varphi}\right)^{n-1}\nonumber \\
 & =\left(B\rho\right)\sum_{n=1}^{\infty}\left(ia_{n}\left(s\right)+b_{n}\left(s\right)\right)\left(x+iy\right)^{n-1}
\end{align}
a Fourier Expansion of the field. The Multipole Coefficients $a_{n},b_{n}$
are energy independent.

The corresponding vector potential is obtained from (Poincaré gauge,
$\bar{r}\cdot\bar{A}=0$) \cite{Jackson}
\begin{equation}
\bar{A}\left(\bar{r},t\right)=-\bar{r}\times\intop_{0}^{1}\bar{B}\left(u\bar{r},t\right)udu,\qquad\phi=-\bar{r}\cdot\intop_{0}^{1}\bar{B}\left(u\bar{r},t\right)udu
\end{equation}
which gives
\begin{align}
\frac{q}{p_{0}}A_{s}\left(s\right) & =\mathrm{Re}\sum_{n=1}^{\infty}\frac{1}{n}\left(ia_{n}\left(s\right)+b_{n}\left(s\right)\right)\left(re^{i\varphi}\right)^{n}\nonumber \\
 & =\mathrm{Re}\sum_{n=1}^{\infty}\frac{1}{n}\left(ia_{n}\left(s\right)+b_{n}\left(s\right)\right)\left(x+iy\right)^{n}.\label{eq:Mult_Exp}
\end{align}
The quadratic Hamiltonian, i.e., ``Potential Function'' (the Total
Energy) for the linear equations of motion, is for mid-plane symmetry
($\bar{x}\equiv\left[x,p_{x},y,p_{y},\Delta s,\delta\right]$) \cite{SLS_1997,CERN-88-05}
\begin{equation}
H\left(\bar{x};s\right)=H_{1}\left(\bar{x};s\right)+H_{2}\left(\bar{x};s\right)
\end{equation}
where
\begin{equation}
H_{1}\left(\bar{x};s\right)=-\frac{x\delta}{\rho\left(s\right)}
\end{equation}
is the Lie Generator for Linear Horizontal Dispersion vs. the energy
deviation
\begin{equation}
p_{t}\equiv\frac{E-E_{0}}{p_{0}c_{0}}
\end{equation}
whereas 
\begin{equation}
H_{2}\left(\bar{x};s\right)=\frac{p_{x}^{2}+p_{y}^{2}}{2\left(1+p_{t}\right)}+\frac{K_{x}x^{2}-K_{y}y^{2}}{2},\qquad K_{x}\equiv b_{2}\left(s\right)+\frac{1}{\rho^{2}\left(s\right)},\qquad K_{y}\equiv b_{2}\left(s\right)\label{eq:H_2}
\end{equation}
generates the Betatron Motion.

The solution can be expressed as a Lie Series (Transport Matrix) \cite{Grobner_1,Lie}

\begin{equation}
\bar{x}_{1}=M\bar{x}_{0}=e^{L\mathscr{D}\left(-H\right)}\bar{x}_{0}\equiv\sum_{k=0}^{\infty}\frac{\left(L\mathscr{D}\left(-H\right)\right)^{k}}{k!}\bar{x}_{0}\label{eq:Lie_series}
\end{equation}
where $L$ is the element length.

The Lie Series forms a Differential Algebra, i.e., it is closed under
the operations: $\left[+,-,\times,/,\partial\right]$; which preserves
the Symplectic Flow. It was applied to Celestial Mechanics in the
late 1960s, e.g. the (academic) ``Main Problem'' ref. \cite{Deprit_1,Deprit_2}
(the 3-Body Problem: Sun-Earth-Moon in which the leading order is
Hill's Equation \cite{Hill}); as well as the ``Engineering-Science''
Optimal Problem \cite{Grobner_2}:
\begin{quotation}
\textit{Soft Landing on the Moon with Fuel Minimization}
\end{quotation}
The solution for the Quadratic Hamiltonian Eq. \eqref{eq:H_2} is
of the form (Transport Matrices)

\begin{equation}
\bar{x}\left(s\right)=M\left(s\right)\bar{x}_{0}
\end{equation}
where $M\left(s\right)$ is the Transport Matrix from $x_{0}\left(s\right)$
to $x_{1}\left(s\right)$.

N.B.: While the equations of motion for a piece-wise constant potential
have straightforward trigonometric solutions, the Lie Series approach
is more practical because it provides for a straightforward generalization
to the nonlinear case \cite{SLS_1997}.

Similar expressions hold for the vertical \& longitudinal planes.

\subsection{Linear Dispersion \& Dispersion Action}

The Linear Horizontal Dispersion is
\begin{equation}
\bar{\eta}_{1}=e^{\mathscr{D}\left(-H\right)}\bar{\eta}_{0}=M_{01}\bar{\eta}_{0}+D,\qquad\bar{\eta}\equiv\left[\begin{array}{c}
\eta_{x}\\
\eta'_{x}
\end{array}\right]\label{eq:Lin_Disp}
\end{equation}
and the (differential) Geometry of momentum changes is described by
the Linear Dispersion Action-Angle Coordinates 
\begin{equation}
\left\{ \begin{aligned}\mathscr{H}_{x}\left(s\right)\equiv & \tilde{\eta}^{\mathrm{T}}\tilde{\eta}=\left(A\left(s\right)A^{\mathrm{T}}\left(s\right)\right)^{-1}\bar{\eta}=\gamma_{x}\left(s\right)\eta_{x}^{2}\left(s\right)+2\alpha_{x}\left(s\right)\eta_{x}\left(s\right)\eta'_{x}\left(s\right)+\beta_{x}\left(s\right){\eta'\left(s\right)}_{x}^{2}\\
\varphi_{x}\left(s\right)= & -\atan2\left(\frac{\tilde{\eta}'_{x}\left(s\right)}{\tilde{\eta}_{x}\left(s\right)}\right)=-\atan2\left(\frac{\alpha_{x}\left(s\right)\eta_{x}\left(s\right)+\beta_{x}\left(s\right)\eta'_{x}\left(s\right)}{\eta_{x}\left(s\right)}\right)
\end{aligned}
\right.
\end{equation}
where (Floquet Space)
\begin{equation}
\tilde{\eta}\equiv A^{-1}\bar{\eta}
\end{equation}
with
\begin{equation}
A=\left[\begin{array}{cc}
\sqrt{\beta_{x}} & 0\\
-\frac{\alpha_{x}}{\sqrt{\beta_{x}}} & \frac{1}{\sqrt{\beta_{x}}}
\end{array}\right],\qquad A^{-1}=\left[\begin{array}{cc}
\frac{1}{\sqrt{\beta_{x}}} & 0\\
\frac{\alpha_{x}}{\sqrt{\beta_{x}}} & \sqrt{\beta_{x}}
\end{array}\right],
\end{equation}
and
\begin{equation}
AA^{\mathrm{T}}=\left[\begin{array}{cc}
\beta_{x} & -\alpha_{x}\\
-\alpha_{x} & \gamma_{x}
\end{array}\right],\qquad\left(AA^{\mathrm{T}}\right)^{-1}=\left[\begin{array}{cc}
\gamma_{x} & \alpha_{x}\\
\alpha_{x} & \beta_{x}
\end{array}\right]
\end{equation}
see next section for the origin of $A$.

The Linear Dispersion Action $\mathscr{H}_{x}\left(s\right)$ is invariant
for drifts \& quadrupoles
\begin{equation}
\tilde{\eta}_{1}=R\left(\varphi_{x}\right)\tilde{\eta}_{0},\qquad R\left(\varphi_{x}\right)\equiv\left[\begin{array}{cc}
\cos\left(\varphi_{x}\right) & \sin\left(\varphi_{x}\right)\\
-\sin\left(\varphi_{x}\right) & \cos\left(\varphi_{x}\right)
\end{array}\right]
\end{equation}
that generate a Floquet Space rotation, see Appendix \ref{subsec:Dispersion_Action}
for the details.

\subsection{Action-Angle Coordinates}

The betatron motion is modelled by

\begin{equation}
\bar{x}_{1}=e^{L\mathscr{D}\left(-H_{2}\right)}\bar{x}_{0}=M_{01}\bar{x}_{0}
\end{equation}
where $L$ is the element length, and the ansatz (Pseudo-Harmonic
Oscillator) 
\begin{equation}
\bar{x}\left(s\right)=\sqrt{2J_{x}\beta_{x}\left(s\right)}\cos\left(\mu_{x}\left(s\right)+\phi_{x}\right)+\eta_{x}\left(s\right)p_{t}\label{eq:pseudo-harm_osc}
\end{equation}
leads to (Phase Space; solutions are ellipses)
\begin{equation}
\bar{x}\left(s\right)=\left[\begin{array}{c}
\sqrt{2J_{x}\beta_{x}\left(s\right)}\cos\left(\mu_{x}\left(s\right)+\phi_{x}\right)+\eta_{x}\left(s\right)p_{t}\\
-\sqrt{\frac{2J_{x}}{\beta_{x}\left(s\right)}}\left(\sin\left(\mu_{x}\left(s\right)+\phi_{x}\right)+\alpha_{x}\left(s\right)\cos\left(\mu_{x}\left(s\right)+\phi_{x}\right)\right)+\eta'_{x}\left(s\right)p_{t}
\end{array}\right]\label{eq:betatron_motion_1}
\end{equation}
where (Courant \& Snyder Phase Advance)
\begin{equation}
\mu_{x}\left(s\right)\equiv\int_{0}^{s}\frac{1}{\beta_{x}\left(u\right)}\totd u\label{eq:C-S_Phase-Advance}
\end{equation}
and
\begin{equation}
\alpha_{x}\left(s\right)\equiv-\frac{\beta'_{x}\left(s\right)}{2}
\end{equation}
see Appendix \eqref{subsec:Action-Angle_Coord} for the details.

The Transport Matrix is diagonalized by the transformation 
\begin{equation}
M_{01}=A_{1}R\left(\Delta\mu\right)A_{0}^{-1}\label{eq:M_diag}
\end{equation}
where
\begin{equation}
A\left(s\right)=\left[\begin{array}{cc}
\sqrt{\beta_{x}\left(s\right)} & 0\\
-\frac{\alpha_{x}\left(s\right)}{\sqrt{\beta_{x}\left(s\right)}} & \frac{1}{\sqrt{\beta_{x}\left(s\right)}}
\end{array}\right],\qquad A^{-1}\left(s\right)=\left[\begin{array}{cc}
\frac{1}{\sqrt{\beta_{x}\left(s\right)}} & 0\\
\frac{\alpha_{x}\left(s\right)}{\sqrt{\beta_{x}\left(s\right)}} & \sqrt{\beta_{x}\left(s\right)}
\end{array}\right]\label{eq:A_mat}
\end{equation}
i.e., with the particular choice $a_{12}=0$, originating from the
ansatz Eq. \eqref{eq:pseudo-harm_osc} and \eqref{eq:C-S_Phase-Advance};
since without it the transformation $A_{1},A_{0}^{-1}$ is not unique.

Hence, one may introduce the State-Space (Floquet Space; solutions
are circles, i.e., Harmonic Oscillator)
\begin{equation}
\tilde{x}\left(s\right)=A^{-1}\left(\bar{x}\left(s\right)-\bar{\eta}\left(s\right)\delta\right)=\left(\Omega^{-1}A^{\mathrm{T}}\left(s\right)\Omega\right)\bar{x}\left(s\right)-D^{-1}\left(s\right)\delta=\left[\begin{array}{c}
\sqrt{2J_{x}}\cos\left(\mu_{x}\left(s\right)+\phi_{x}\right)\\
-\sqrt{2J_{x}}\sin\left(\mu_{x}\left(s\right)+\phi_{x}\right)
\end{array}\right]
\end{equation}
see Fig. \eqref{fig:Phase-Space_Fl-Space}.

The Action-Angle Coordinates $\left[\phi_{x},J_{x}\right]$, i.e.,
the Invariants for the System, are \cite{C-S}
\begin{equation}
\left\{ \begin{aligned}2J_{x}= & \tilde{x}^{\mathrm{T}}\tilde{x}=\left(A^{-1}\bar{x}\right)^{\mathrm{T}}A^{-1}\bar{x}=\left(\Omega\bar{x}\right)^{\mathrm{T}}AA^{\mathrm{T}}\left(\Omega\bar{x}\right)=\gamma_{x}\left(s\right)x^{2}+2\alpha_{x}\left(s\right)xp_{x}\left(s\right)+\beta_{x}p_{x}^{2}\left(s\right)\\
\phi_{x}= & -\atan2\left(\frac{\tilde{p}_{x}}{\tilde{x}}\right)=-\atan2\left(\frac{\alpha_{x}\left(s\right)x+\beta_{x}p_{x}\left(s\right)}{x}\right)
\end{aligned}
\right.\label{eq:action_angle_var}
\end{equation}
and the corresponding Ellipse Parameters are summarized in Fig. \ref{fig:Phase-Space_Ellips}.
For the details see Appendix \ref{subsec:Action-Angle_Coord}.
\begin{center}
\begin{figure}[H]
\centering{}\includegraphics[width=10cm]{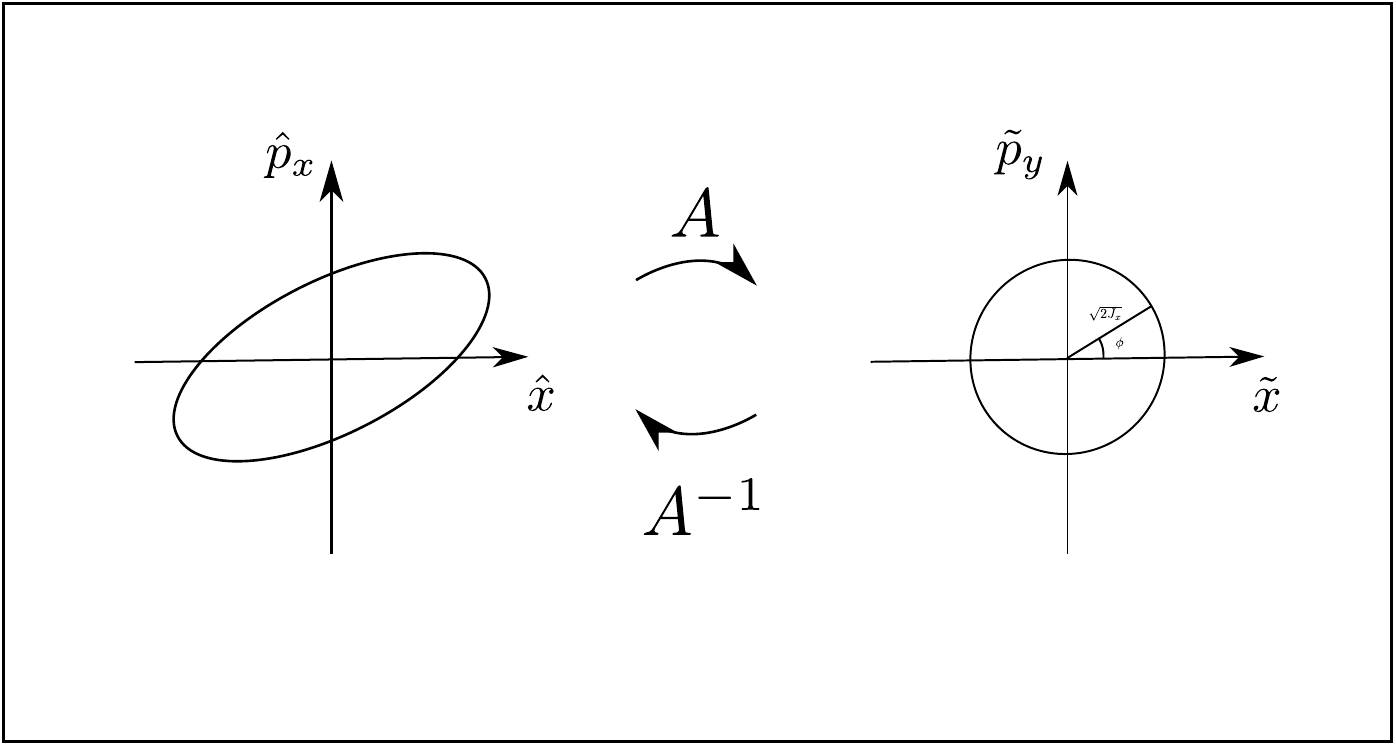}\caption{Phase-Space and Floquet Space.\label{fig:Phase-Space_Fl-Space}}
\end{figure}
\par\end{center}

\begin{center}
\begin{figure}[H]
\centering{}\includegraphics[width=9cm]{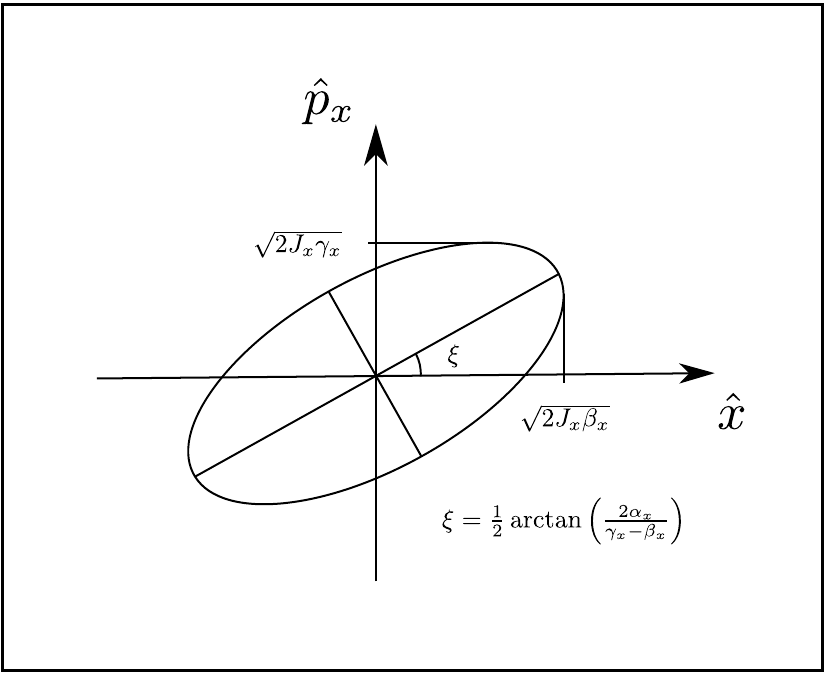}\caption{Phase Space Geometry.\label{fig:Phase-Space_Ellips}}
\end{figure}
\par\end{center}

\subsection{Linear Optics}

For convenience, one may generally use
\begin{equation}
\sigma\equiv AA^{\mathrm{T}}=\left[\begin{array}{cc}
\beta_{x} & -\alpha_{x}\\
-\alpha_{x} & \gamma_{x}
\end{array}\right],\qquad\gamma_{x}\equiv\frac{1+\alpha_{x}^{2}}{\beta_{x}}\label{eq:A_At_1}
\end{equation}
for the symmetric matrix generating the Quadratic Form for the action.
It is propagated by (bilinear transformation)
\begin{equation}
\sigma_{k}=A_{k}A_{k}^{\mathrm{T}}=M_{j\rightarrow k}A_{j}\left(M_{j\rightarrow k}A_{j}\right)^{\mathrm{T}}=M_{j\rightarrow k}\sigma_{j}M_{j\rightarrow k}^{\mathrm{T}}\label{eq:A_At_2}
\end{equation}
where $M_{j\rightarrow k}$ is the Transport Matrix from $j\rightarrow k$.

N.B. For numerical calculations of the Linear Optics, a more streamlined
approach is to introduce
\begin{align}
A_{k}\left(\Delta\mu_{x,k}\right)\equiv A_{k}R\left(\Delta\mu_{x,k}\right) & =\left[\begin{array}{cc}
m_{11} & m_{12}\\
m_{21} & m_{22}
\end{array}\right]=\left[\begin{array}{cc}
\sqrt{\beta_{x}\left(s\right)} & 0\\
-\frac{\alpha_{x}\left(s\right)}{\sqrt{\beta_{x}\left(s\right)}} & \frac{1}{\sqrt{\beta_{x}\left(s\right)}}
\end{array}\right]\left[\begin{array}{cc}
\cos\left(\Delta\mu_{x,k}\right) & \sin\left(\Delta\mu_{x,k}\right)\\
-\sin\left(\Delta\mu_{x,k}\right) & \cos\left(\Delta\mu_{x,k}\right)
\end{array}\right]\nonumber \\
 & =\left[\begin{array}{cc}
\sqrt{\beta_{x,k}}\cos\left(\Delta\mu_{x,k}\right) & \sqrt{\beta_{x,k}}\sin\left(\Delta\mu_{x,k}\right)\\
-\frac{\sin\left(\Delta\mu_{x,k}\right)+\alpha_{x,k}\cos\left(\Delta\mu_{x,k}\right)}{\sqrt{\beta_{x,k}}} & -\frac{\alpha_{x,k}\sin\left(\Delta\mu_{x,k}\right)-\cos\left(\Delta\mu_{x,k}\right)}{\sqrt{\beta_{x,k}}}
\end{array}\right]
\end{align}
and instead compute
\begin{equation}
\begin{cases}
A_{k}\left(\Delta\mu_{x,k}\right) & =M_{j\rightarrow k}A_{j}\left(\Delta\mu_{x,j}\right)\\
\alpha_{x,k} & =-m_{11}m_{21}-m_{12}m_{22}\\
\beta_{x,k} & =m_{11}^{2}+m_{12}^{2}\\
\Delta\mu_{x,k} & =\atan2\left(\frac{m_{12}}{m_{11}}\right)
\end{cases}\label{eq:A_prop}
\end{equation}
 where Eq. \eqref{eq:M_diag} has been used. The inverse transformation
is
\begin{equation}
A_{k}\equiv A_{k}\left(0\right)=R^{-1}\left(\Delta\mu_{x,k}\right)A_{j}\left(\Delta\mu_{x,k}\right)
\end{equation}
where
\begin{equation}
R^{-1}\left(\Delta\mu\right)\equiv\left[\begin{array}{cc}
\cos\left(\Delta\mu\right) & -\sin\left(\Delta\mu\right)\\
\sin\left(\Delta\mu\right) & \cos\left(\Delta\mu\right)
\end{array}\right].
\end{equation}

The parametrized Transport Matrix and Dispersion Vector is
\begin{align}
\left.\left(M+D\right)\right|_{\delta=0} & =\left[\begin{array}{cc}
m_{11} & m_{12}\\
m_{21} & m_{22}
\end{array}\right]+\left[\begin{array}{c}
m_{16}\\
m_{26}
\end{array}\right]\nonumber \\
 & =\left[\begin{array}{cc}
\sqrt{\frac{\beta_{1,x}}{\beta_{0,x}}}\left(\cos\left(\Delta\mu_{x}\right)+\alpha_{0,x}\sin\left(\Delta\mu_{x}\right)\right) & \sqrt{\beta_{0,x}\beta_{1,x}}\sin\left(\Delta\mu_{x}\right)\\
-\frac{\left(1+\alpha_{0,x}\alpha_{1,x}\right)\sin\left(\Delta\mu_{x}\right)+\left(\alpha_{1,x}-\alpha_{0,x}\right)\cos\left(\Delta\mu_{x}\right)}{\sqrt{\beta_{0,x}\beta_{1,x}}} & \sqrt{\frac{\beta_{0,x}}{\beta_{1,x}}}\left(\cos\left(\Delta\mu_{x}\right)-\alpha_{1,x}\sin\left(\Delta\mu_{x}\right)\right)
\end{array}\right]+\left[\begin{array}{c}
m_{16}\\
m_{26}
\end{array}\right]
\end{align}
where
\begin{equation}
\Delta\mu_{x}=\atan2\frac{m_{12}}{\beta_{0,x}m_{11}-\alpha_{0,x}m_{12}}=\atan2\frac{m_{12}}{\beta_{1,x}m_{22}+\alpha_{1,x}m_{12}}
\end{equation}
with the convention (Linear Vector Space)
\begin{equation}
\bar{x}\rightarrow M\bar{x}+Dp_{t}.
\end{equation}

\subsection{Periodic System}

For a Periodic System it simplifies to (Poincaré Map)
\begin{equation}
M+D=\left[\begin{array}{cc}
\cos\left(2\pi\nu_{x}\right)+\alpha\sin\left(2\pi\nu_{x}\right) & \beta_{x}\sin\left(2\pi\nu_{x}\right)\\
-\gamma_{x}\sin\left(2\pi\nu_{x}\right) & \cos\left(2\pi\nu_{x}\right)-\alpha_{x}\sin\left(2\pi\nu_{x}\right)
\end{array}\right]+\left[\begin{array}{c}
m_{16}\\
m_{26}
\end{array}\right]\label{eq:per_map}
\end{equation}
which can be Diagonalized (Floquet Space)
\begin{equation}
M=AR\left(2\pi\nu_{x}\right)A^{-1},\qquad\tilde{D}\equiv A^{-1}\left[\begin{array}{c}
m_{16}\\
m_{26}
\end{array}\right]\label{eq:per_map_2}
\end{equation}
with the Linear Dispersion
\begin{equation}
\bar{\eta}=M\bar{\eta}+D\Rightarrow\bar{\eta}=\left(I-M\right)^{-1}D\label{eq:dispersion}
\end{equation}
and Tune
\begin{equation}
\nu_{x}\left(\delta\right)\equiv\frac{\mu_{x}\left(p_{t}\right)}{2\pi}=\nu_{x,0}+\frac{1}{2\pi}\left(\xi_{x}^{\left(1\right)}p_{t}+\xi_{x}^{\left(2\right)}p_{t}^{2}+\ldots\right)=\frac{\arccos\left(\frac{\mathrm{Tr}\left\{ M\left(p_{t}\right)\right\} }{2}\right)}{2\pi}\label{eq:tune}
\end{equation}
where $\xi_{x}^{\left(1\right)}$ is the Linear Chromaticity, $\xi_{x}^{\left(2\right)}$
is the 2nd Order Chromaticity, etc.; global properties of the cell
(\& lattice).

Similar expressions hold for the vertical plane.\newpage{}

\section{A Use Case Approach\label{sec:Use-Case_Approach}}

Perhaps, the lack of effective, interactive CAD Tools for e.g. Beam
Dynamics Modelling \& Lattice design for Particle Accelerators \textendash{}
something that has been available to e.g. Mechanical Engineers since
the mid-1960s \textendash{} is a reflection of poorly architectured
software \& infrastructure for the underlying Beam Dynamics Models.
Clearly, effective numerical algorithms for ditto are known; i.e.,
papers have been written about it since the late 1980s. One symptom
of this, is that the ``Open Source'' approach \cite{GNU} is not
standard practice. However, exceptions exists, e.g. \cite{Use_Case_1,FLAME,Tracy-2_SLS_2,Tracy-2_AT,EPICS,MML-TK,MML,pyAT_1,pyAT_2,Will_R}.

Contrarily, the \textquotedblleft Use-Case\textquotedblright{} approach
is a strategy for System Engineering pursued by e.g. the Software
Industry to capture the often elusive functional specs. from often
vague, unknown, and changing requirements for the End-Users of a complex
system (i.e. ``fluid work environment''). In particular, an intuitive
approach where the Users and the System are visualized as a set of
\textquotedblleft Actors\textquotedblright{} (End-User or other System)
interacting with a \textquotedblleft Black-Box\textquotedblright ,
see Fig. \ref{fig:Use-Case_Approach}, Fig. 1 in ref. \cite{Use_Case_3}.
\begin{center}
\begin{figure}[H]
\centering{}\includegraphics[width=8cm]{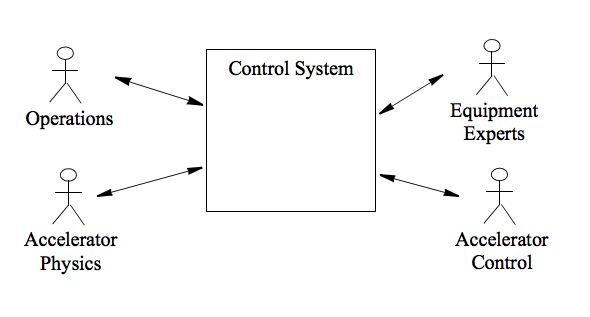}\caption{``Use Case'' Approach.\label{fig:Use-Case_Approach}}
\end{figure}
\par\end{center}

As is often the case though, what today have become ``Best Practices''
for the design \& implementation of software \& hardware architectures
for complex telecomm systems, originated from the solution of an intricate
commercial Systems Engineering Problem in the 1980s. To quote I. Jacobson,
2010, one of the \textquotedblleft \textit{Three Amigos}\textquotedblright{}
who with G. Booch and R. Rambaugh invented \& developed the Unified
Modeling Language (UML) for Software Engineering in the 1990s \cite{UML,Booch,Jacobson}:
\begin{quotation}
\textit{It was like that in the late 1960s and the \textquoteleft 70s
when the Ericsson AXE system beat all competition and won every contract
thanks to being component-based. Similarly, when Rational was successful
because of UML and Objectory. And Telelogic because of SDL.}\footnote{Specification and Description Language for Distributed Reactive Systems
\cite{SDL}. The system is specified by

Extended Finite State Machines: Interconnected Abstract Machines.
The language is formally complete (in

the sense of Gödel \cite{Godel_1,Found_of_Math,Godel_2}), i.e., can
automatically generate the computer code for the system.}

\textit{...}

\textit{In 1986, Use Case was the solution to the problem that traditional
functional specifications were immense and not testable. To start
from the users and find their different use cases made the specifications
understandable, while we also at the same time found the test cases.
The result was a good way to do test-driven development, now being
popular in agile teams.}
\end{quotation}
Also, as G. Booch wrote in the foreword to \textquotedblleft Design
Patterns: Elements of Reusable Object-Oriented Software\textquotedblright ,
1994 \cite{Design_Patterns}:
\begin{quotation}
\textit{All well-structured object-oriented architectures are full
of patterns. Indeed, one of the ways that I measure the quality of
an object-oriented system is to judge whether or not its developers
have paid careful attention to the common collaborations among its
objects. Focusing on such mechanisms during a system\textquoteright s
development can yield an architecture that is smaller, simpler, and
far more understandable than if these patterns are ignored.}

\textit{The importance of patterns in crafting complex systems has
been long recognized in other disciplines. In particular, Christopher
Alexander (\cite{C_Alexander}) and his colleagues were perhaps the
first to propose the idea of using a pattern language to architect
buildings and cities. His ideas and the contributions of others have
now taken root in the object-oriented software community. In short,
the concept of the design pattern provides a key to helping developers
leverage the expertise of other skilled architects.}

...
\end{quotation}
which was written by the \textquotedblleft \textit{Gang of Four}\textquotedblright :
E. Gamma, R. Helm, R. Johnson, and J. Vlissides. Alexander's work
also inspired the implementation of the first Wiki by Cunningham \cite{Wiki_0.0}.

Not surprisingly, to start from the End-User is part of the strategy
for systematic database design as well. To quote T. Halpin\footnote{Halpin formalized and introduced \textquotedblleft Object-Role Modeling\textquotedblright{}
(ORM) with his 1989 thesis \cite{Halpin_2}. A generalization of the
Semantic Modeling of Information Systems in Europe in the 1970s; based
on a Graph Model rather than e.g. Relational or Hierarchical Models
\cite{Halpin_3}. Roughly, \textquotedblleft Relationships\textquotedblright{}
are generalized to \textquotedblleft Roles\textquotedblright .} \cite{Halpin_1}:
\begin{quotation}
\textit{Although a rigorous process model is best built on top of
a data model, an overview of the processes can be useful as a precursor
to the data modeling, especially if the application is large or only
vaguely understood. It is often helpful to get a clear picture of
the functions of the application first.}
\end{quotation}
Similarly, the field of Particle Accelerators has started to catch
on \cite{Use_Case_1,Tracy-2_SLS_2,Tracy-2_DIAMOND_2,Tracy-2_NSLS-II,Tracy-2_Python,Tracy-2_Cython}.

In conclusion, effective, goal oriented guidelines for how to pursue
a systematic End-User oriented approach for the functional specs for
a software architecture of arbitrary complexity; for e.g. model-based
control, end-to-end testing of controls application before commissioning
of the system, etc.

\subsection{Lattice: Independent Parameters}

The linearised beam dynamics model is a straightforward exercise in
Hamiltonian Dynamics and Linear Control Theory.

The Hamiltonian is Eq. \eqref{eq:H_2} ($\bar{x}\equiv\left[x,p_{x},y,p_{y},c_{0}t,-p_{t}\right]$)
\begin{equation}
H_{2}\left(\bar{x};s\right)=\frac{p_{x}^{2}+p_{y}^{2}}{2\left(1+p_{t}\right)}+\frac{b_{2}\left(s\right)}{2}\left(x^{2}-y^{2}\right)+\frac{x^{2}}{2\rho^{2}\left(s\right)}-\frac{xp_{t}}{\rho\left(s\right)}
\end{equation}
i.e., what is defined by the Lattice File. The Equations of Motion
for the horizontal plane are (Hamiltons Equations)
\begin{equation}
\begin{cases}
x'= & \partial_{p_{x}}H=\frac{p_{x}}{1+p_{t}}\\
p'_{x}= & -\partial_{x}H=-b_{2}\left(s\right)x-\frac{x}{\rho^{2}\left(s\right)}+\frac{p_{t}}{\rho\left(s\right)}\\
y'= & \partial_{p_{y}}H=\frac{p_{y}}{1+p_{t}}\\
p'_{y}= & -\partial_{y}H=b_{2}\left(s\right)y
\end{cases}
\end{equation}
By taking the derivative
\begin{equation}
x''=\frac{p'_{x}}{1+p_{t}},\qquad y''=\frac{p'_{y}}{1+p_{t}}
\end{equation}
the system of first order Ordinary Differential Equations (ODEs) can
be reduced to a system of second order ODEs (Hill's Equation \cite{Hill})
\begin{equation}
\begin{cases}
x''+\frac{K_{x}\left(s\right)}{1+p_{t}}x & =\frac{\delta}{\rho\left(s\right)\left(1+p_{t}\right)}\\
y''-\frac{K_{y}\left(s\right)}{1+p_{t}}y & =0
\end{cases}
\end{equation}
where
\begin{equation}
K_{x}\left(s\right)=b_{2}\left(s\right)-\frac{1}{\rho^{2}\left(s\right)},\qquad K_{y}\left(s\right)=-b_{2}\left(s\right).
\end{equation}
The Lattice Parameters, typically piece-wise constant, i.e., independent,
are summarized in Tab. \ref{tab:Lat_Param}.

The total energy is given by Eq. (\ref{eq:rel_energy}):
\begin{equation}
E_{\mathrm{b}}=\sqrt{\left(m_{0}c_{0}^{2}\right)^{2}+\left(pc_{0}\right)^{2}}.
\end{equation}

\begin{table}[H]
\centering{}%
\begin{tabular}{|c|c|c|}
\hline 
Parameter Name & Symbol & Units\tabularnewline
\hline 
\hline 
Beam Energy & $E_{\mathrm{b}}$ & $\left[\mathrm{GeV}\right]$\tabularnewline
\hline 
Element Length & $L\left(s\right)$ & $\left[\mathrm{m}\right]$\tabularnewline
\hline 
Dipole Bend Radius & $\rho\left(s\right)=\frac{L}{\phi}$ & $\left[\mathrm{m}\right]$\tabularnewline
\hline 
Quadrupole Gradient & $b_{2}\left(s\right)$ & $\left[\mathrm{m}^{-2}\right]$\tabularnewline
\hline 
\end{tabular}\caption{Lattice Parameters (independent).\label{tab:Lat_Param}}
\end{table}

\subsection{Linear Optics: Dependent Parameters}

The solution for a piece-wise constant $K\left(s\right)$ is the Transport
Matrix obtained in Appendix \ref{subsec:Action-Angle_Coord}
\begin{equation}
\bar{x}_{j\rightarrow k}=M_{j\rightarrow k}\bar{x}_{0}
\end{equation}
where $M_{j\rightarrow k}$ is the Transport Matrix from $j\rightarrow k$.

The Linear Dispersion \& Optics (Twiss Parameters \cite{C-S,Twiss_Frank})
are propagated by Eqs. \eqref{eq:Lin_Disp} \& \eqref{eq:A_At_2}
\begin{align}
\bar{\eta}_{k} & =M_{j\rightarrow k}\bar{\eta}_{j},\nonumber \\
A_{k}\left(\overline{\Delta\mu}_{k}\right) & =M_{j\rightarrow k}A_{j}\left(\overline{\Delta\mu}_{j}\right),\nonumber \\
\alpha_{k,\left[x,y\right]} & =-m_{11}m_{21}-m_{12}m_{22},\nonumber \\
\beta_{k,\left[x,y\right]} & =m_{11}^{2}+m_{12}^{2}
\end{align}
 where
\begin{equation}
A_{k}\left(\overline{\Delta\mu}_{k}\right)\equiv\left[\begin{array}{ccc}
A_{k}\left(\Delta\mu_{x,k}\right) & 0 & 0\\
0 & A_{k}\left(\Delta\mu_{y,k}\right) & 0\\
0 & 0 & 1
\end{array}\right]
\end{equation}
with the block diagonal sub matrices
\begin{equation}
A_{k}\left(\Delta\mu_{\left[x,y\right],k}\right)\equiv\left[\begin{array}{cc}
m_{11} & m_{12}\\
m_{21} & m_{22}
\end{array}\right]=\left[\begin{array}{cc}
\sqrt{\beta_{k,\left[x,y\right]}}\cos\left(\Delta\mu_{k,\left[x,y\right]}\right) & \sqrt{\beta_{k,\left[x,y\right]}}\sin\left(\Delta\mu_{k,\left[x,y\right]}\right)\\
-\frac{\sin\left(\Delta\mu_{k,\left[x,y\right]}\right)+\alpha_{k,\left[x,y\right]}\cos\left(\Delta\mu_{k,\left[x,y\right]}\right)}{\sqrt{\beta_{k,\left[x,y\right]}}} & \frac{\cos\left(\Delta\mu_{k,\left[x,y\right]}\right)-\alpha_{x,k}\sin\left(\Delta\mu_{k,\left[x,y\right]}\right)}{\sqrt{\beta_{k,\left[x,y\right]}}}
\end{array}\right].
\end{equation}

Similarly, for a Periodic System the Linear Dispersion \& Optics Functions
are given by Eqs. \eqref{eq:A_prop}, \eqref{eq:per_map_2}, and \eqref{eq:dispersion}
\begin{align}
\bar{\eta} & =\left(I-M\right)^{-1}D,\nonumber \\
M & =A_{0}R\left(2\pi\bar{\nu}\right)A_{0}^{-1},\nonumber \\
A_{k}\left(\overline{\Delta\mu}_{k}\right) & =M_{j\rightarrow k}A_{j}\left(\overline{\Delta\mu}_{j}\right),\nonumber \\
\alpha_{\left[x,y\right],k} & =-m_{11}m_{21}-m_{12}m_{22},\nonumber \\
\beta_{\left[x,y\right],k} & =m_{11}^{2}+m_{12}^{2}
\end{align}
where
\begin{equation}
D=\left[m_{16},m_{26},0,0,0,0\right]^{\mathrm{T}},
\end{equation}
the 2nd equation a Matrix Diagonalization with
\begin{equation}
R\left(2\pi\bar{\nu}\right)\equiv\left[\begin{array}{ccc}
R\left(\Delta\mu_{x,k}\right) & 0 & 0\\
0 & R\left(\Delta\mu_{y,k}\right) & 0\\
0 & 0 & 1
\end{array}\right],\qquad R\left(\Delta\mu\right)\equiv\left[\begin{array}{cc}
\cos\left(\Delta\mu\right) & \sin\left(\Delta\mu\right)\\
-\sin\left(\Delta\mu\right) & \cos\left(\Delta\mu\right)
\end{array}\right],
\end{equation}
and the 3rd equation transports the Linear Optics Functions from the
beginning of the Lattice.

The Local Linear Optics design parameters, i.e., dependent parameters,
along the structure are:

\begin{table}[H]
\centering{}%
\begin{tabular}{|c|c|c|}
\hline 
Parameter Name & Symbol & Units\tabularnewline
\hline 
\hline 
Hor. Linear Dispersion & $\bar{\eta}\left(s\right)\equiv\left[\eta_{x}\left(s\right),\eta'_{x}\left(s\right)\right]$ & $\left[\mathrm{m}\right]$\tabularnewline
\hline 
Beta Function & $\bar{\beta}\left(s\right)\equiv\left[\beta_{x}\left(s\right),\beta_{y}\left(s\right)\right]$ & $\left[\mathrm{m}\right]$\tabularnewline
\hline 
Beta Function Derivative & $\bar{\alpha}\left(s\right)\equiv-\frac{1}{2}\partial_{s}\bar{\beta}\left(s\right)=\left[\alpha_{x}\left(s\right),\alpha_{y}\left(s\right)\right]$ & $\left[.\right]$\tabularnewline
\hline 
Normalized Phase Advance & $\bar{\nu}\left(s\right)\equiv\frac{\bar{\mu}\left(s\right)}{2\pi}=\left[\frac{\mu_{x}\left(s\right)}{2\pi},\frac{\mu_{y}\left(s\right)}{2\pi}\right]$ & $\left[.\right]$\tabularnewline
\hline 
\end{tabular}\caption{Local Linear Optics Design Parameters (dependent).\label{tab:Local_Lin_Optics}}
\end{table}

And the Global Linear Optics for a Periodic Cell are:

\begin{table}[H]
\centering{}%
\begin{tabular}{|c|c|c|}
\hline 
Parameter Name & Symbol & Units\tabularnewline
\hline 
\hline 
Cell Tune & $\bar{\nu}\equiv\left[\frac{\mu_{x}\left(L_{\mathrm{tot}}\right)}{2\pi},\frac{\mu_{y}\left(L_{\mathrm{tot}}\right)}{2\pi}\right]$ & $\left[\mathrm{.}\right]$\tabularnewline
\hline 
Linear Chromaticity & $\bar{\xi}\equiv\left[\xi_{x},\xi_{y}\right]$ & $\left[\mathrm{.}\right]$\tabularnewline
\hline 
Horizontal Emittance & $\varepsilon_{x}$ & $\left[\mathrm{pm\cdot rad}\right]$\tabularnewline
\hline 
Linear Dispersion Action & $\mathscr{H}_{x}$ & $\left[\mathrm{m}\right]$\tabularnewline
\hline 
Momentum Spread & $\sigma_{\delta}$ & $\left[.\right]$\tabularnewline
\hline 
Linear Momentum Compaction & $\alpha_{\mathrm{c}}$ & $\left[\mathrm{.}\right]$\tabularnewline
\hline 
Energy Loss per Turn & $U_{0}$ & $\left[\mathrm{eV}\right]$\tabularnewline
\hline 
Damping Times & $\bar{\tau}\equiv\left[\tau_{x},\tau_{y},\tau_{z}\right]$ & $\left[\mathrm{msec}\right]$\tabularnewline
\hline 
Damping Partition Numbers & $\bar{J}\equiv\left[J_{x},J_{y},J_{z}\right]$ & $\left[.\right]$\tabularnewline
\hline 
Total Length & $L_{\mathrm{tot}}$ & $\left[\mathrm{m}\right]$\tabularnewline
\hline 
Total Bend Angle & $\phi_{\mathrm{tot}}$ & $\left[\lyxmathsym{\textdegree}\right]$\tabularnewline
\hline 
Total Absolute Bend Angle & $\sum\left|\phi\right|$ & $\left[\lyxmathsym{\textdegree}\right]$\tabularnewline
\hline 
\end{tabular}\caption{Global Linear Optics Properties for a Periodic Cell (dependent).\label{tab:Global_Lin_Optics}}
\end{table}

\subsection{Optimal Problems}

\subsubsection{Least-Square Fit}

The Merit Function for a Least-Square fit has the general form
\begin{equation}
\mathrm{Min}\left\{ \chi^{2}\right\} =\sum_{k}c_{k}\left(y_{k}-y_{k}^{*}\left(x_{k};\bar{a}\right)\right)^{2}
\end{equation}
where $\bar{x}$ are the initial conditions, $\bar{y}^{*},\bar{y}$
the resulting propagated and desired values, respectively, (e.g. for
Linear Dispersion \& Optics, see Tab. \ref{tab:Local_Lin_Optics}-\ref{tab:Global_Lin_Optics})
$\bar{a}$ the independent parameters e.g. Lattice, see Tab. \ref{tab:Lat_Param},
and $\bar{c}$ arbitrary weight factors (since e.g. the units for
the dependent parameters might be different). It can be solved numerically
by using e.g. Powell's Method \cite{Powell,Fletcher-Powell}, Levenberg\textendash Marquardt
\cite{Marquardt,Levenberg}, or Downhill Simplex \cite{Downhill_Simplex}.

\subsubsection{Use Case: Beamline Matching\label{subsec:Matching}}

To Match a Beamline:
\begin{enumerate}
\item The Linear Optics Functions at the Cell Entrance $\bar{x}$ are given.
\item The desired values for the Linear Optics Functions at the Cell Exit
$\bar{y}$ are given.
\item Suitable numerical weight coefficients $c_{k}$ are given; essentially,
to numerically balance terms with different units \& magnitude and
relative importance.
\item The select Lattice Parameters $\bar{a}$ are set to the Initial Values
$\bar{a}_{0}$.
\item The Linear Optics Functions at the Exit $\bar{y}\left(\bar{x};\bar{a}\right)$
are computed, for the given $\bar{x},\bar{a}_{0}$
\begin{equation}
\begin{cases}
\bar{\eta}_{k} & =M_{j\rightarrow k}\bar{\eta}_{j}\\
A_{k}\left(\bar{\Delta\mu}_{k}\right) & =M_{j\rightarrow k}A_{j}\left(\bar{\Delta\mu}_{j}\right)\\
\alpha_{\left[x,y\right],k} & =-m_{11}m_{21}-m_{12}m_{22}\\
\beta_{\left[x,y\right],k} & =m_{11}^{2}+m_{12}^{2}
\end{cases}
\end{equation}
\item The (merit) $\chi^{2}$-function is evaluated; i.e., a measure for
how much the computed values $\bar{y}^{*}\left(\bar{x};\bar{a}\right)$
deviate from the desired $\bar{y}$.
\item The Lattice Parameters $\bar{a}$ are varied to minimize $\chi^{2}$.
\item Repeat until within desired precision or maximum number of iterations.
\end{enumerate}
A summary is given in Tab. \ref{tab:Matching}. For an example (of
a MATLAB implementation based on ``Accelerator Toolbox'') see ref.
\cite{atmatch}.

\begin{table}[H]
\centering{}%
\begin{tabular}{|>{\centering}m{2.5cm}|>{\centering}m{2cm}|>{\centering}m{2cm}|>{\centering}m{2cm}|>{\centering}m{2cm}|>{\centering}m{2.2cm}|c|}
\hline 
Parameter Name & Initial Value

$\bar{a}_{0}$ & Final Value

$\bar{a}_{1}$ & Initial Value

$\bar{x}$ & Final Value

$\bar{y}^{*}$ & Desired Value

$\bar{y}$ & Delta\tabularnewline
\hline 
\hline 
Element Length & $\bar{L}_{0}$ & $\bar{L}_{1}$ &  &  &  & \tabularnewline
\hline 
Bend Radius & $\bar{\rho}_{0}$ & $\bar{\rho}_{1}$ &  &  &  & \tabularnewline
\hline 
Gradient & $\bar{b}_{2,0}$ & $\bar{b}_{2,1}$ &  &  &  & \tabularnewline
\hline 
Dispersion &  &  & $\bar{\eta}_{0}$ & $\bar{\eta}_{1}^{*}$ & $\bar{\eta}_{1}$ & $\bar{\eta}_{1}^{*}-\bar{\eta}_{1}$\tabularnewline
\hline 
Dispersion Derivative &  &  & $\bar{\eta}'_{0}$ & $\bar{\eta}{}_{1}^{'*}$ & $\bar{\eta}'_{1}$ & $\bar{\eta}{}_{1}^{'*}-\bar{\eta}'_{1}$\tabularnewline
\hline 
Beta Function &  &  & $\bar{\beta}_{0}$ & $\bar{\beta}_{1}^{*}$ & $\bar{\beta}_{1}$ & $\bar{\beta}_{1}^{*}-\bar{\beta}_{1}$\tabularnewline
\hline 
Beta Function

Derivative &  &  & $\bar{\alpha}_{0}$ & $\bar{\alpha}_{1}^{*}$ & $\bar{\alpha}_{1}$ & $\bar{\alpha}_{1}^{*}-\bar{\alpha}_{1}$\tabularnewline
\hline 
\end{tabular}\caption{Matching: Independent and Dependent Parameters.\label{tab:Matching}}
\end{table}

\begin{figure}[H]
\centering{}\includegraphics[width=8cm]{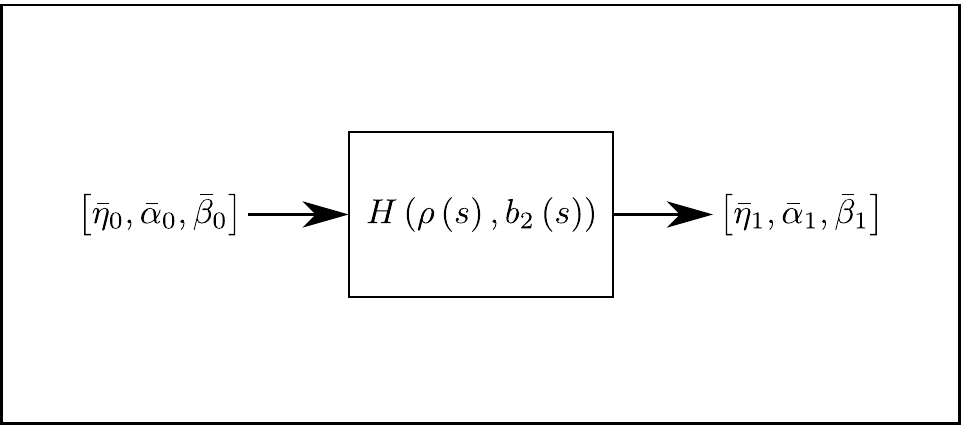}\caption{Matching.\label{fig:Matching}}
\end{figure}

\subsubsection{Use Case: Periodic Cell\label{subsec:Periodic_Cell}}

To optimize a Periodic Solution is akin to Beamline Matching, but
the computation is done for the periodic solution with periodic constraints.
\begin{enumerate}
\item Suitable numerical weight coefficients $c_{k}$ are given; essentially,
to numerically balance terms with different units \& magnitude and
relative importance.
\item The select Lattice Parameters $\bar{a}$ are set to the Initial Values
$\bar{a}_{0}$.
\item The Periodic Linear Optics Functions $\bar{y}\left(\bar{x};\bar{a}\right)$
are computed, for the given $\bar{x},\bar{a}_{0}$
\begin{equation}
\begin{cases}
\bar{\eta} & =\left(I-M\right)^{-1}D\\
M & =A_{0}R\left(2\pi\bar{\nu}\right)A_{0}^{-1}\\
A_{k}\left(\bar{\Delta\mu}_{k}\right) & =M_{j\rightarrow k}A_{j}\left(\bar{\Delta\mu}_{j}\right)\\
\beta_{k,\left[x,y\right]} & =m_{11}^{2}+m_{12}^{2}\\
\varepsilon_{x} & =C_{\mathrm{q}}\gamma^{2}\frac{I_{5}}{I_{2}-I_{4}}
\end{cases}
\end{equation}
where the synchtrotron integrals $I_{2},I_{4},I_{5}$ are given by
Eqs. \eqref{eq:synchr_int}.
\item The (merit) $\chi^{2}$-function is evaluated; i.e., a measure for
how much the computed values $\bar{y}^{*}\left(\bar{x};\bar{a}\right)$
deviate from the desired $\bar{y}$.
\item The Lattice Parameters $\bar{a}$ are varied to minimize $\chi^{2}$.
\end{enumerate}
A summary is given in Tab. \ref{tab:Periodic_Cell}. For an example
(of a MATLAB implementation based on ``Accelerator Toolbox'') see
ref. \cite{atmatch}.

\begin{table}[H]
\centering{}%
\begin{tabular}{|>{\centering}m{3cm}|>{\centering}m{2cm}|>{\centering}m{1.9cm}|>{\centering}m{2.6cm}|>{\centering}m{2.2cm}|>{\centering}m{1.3cm}|}
\hline 
Parameter Name & Initial Value

$\bar{a}_{0}$ & Final Value

$\bar{a}_{1}$ & Computed Value

$\bar{y}^{*}$ & Desired Value

$\bar{y}$ & Delta\tabularnewline
\hline 
\hline 
Element Length & $\bar{L}_{0}$ & $\bar{L}_{1}$ &  &  & \tabularnewline
\hline 
Bend Radius & $\bar{\rho}_{0}$ & $\bar{\rho}_{1}$ &  &  & \tabularnewline
\hline 
Gradient & $\bar{b}_{2,0}$ & $\bar{b}_{2,1}$ &  &  & \tabularnewline
\hline 
Dispersion &  &  & $\bar{\eta}^{*}$ & $\bar{\eta}$ & $\bar{\eta}^{*}-\bar{\eta}$\tabularnewline
\hline 
Dispersion Derivative &  &  & $\bar{\eta}'^{*}$ & $\bar{\eta}'$ & $\bar{\eta}'^{*}-\bar{\eta}'$\tabularnewline
\hline 
Beta Function &  &  & $\bar{\beta}^{*}$ & $\bar{\beta}$ & $\bar{\beta}^{*}-\bar{\beta}$\tabularnewline
\hline 
Cell Tune &  &  & $\bar{\nu}^{*}$ & $\bar{\nu}$ & $\bar{\nu}^{*}-\bar{\nu}$\tabularnewline
\hline 
Hor. Emittance &  &  & $\varepsilon_{x}^{*}$ & $\varepsilon_{x}$ & $\varepsilon_{x}^{*}-\varepsilon$\tabularnewline
\hline 
\end{tabular}\caption{Periodic Cell: Independent and Dependent Parameters.\label{tab:Periodic_Cell}}
\end{table}
\begin{figure}[H]
\centering{}\includegraphics[width=8cm]{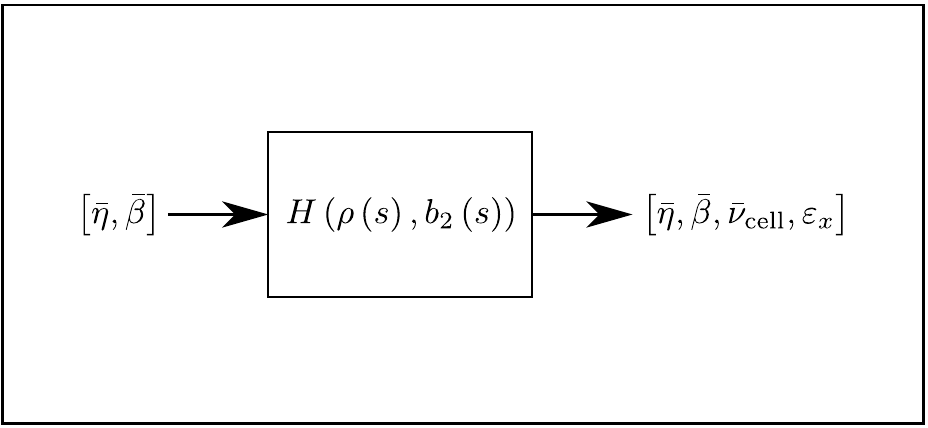}\caption{Periodic Cell.\label{fig:Periodic_Cell.}}
\end{figure}

\subsubsection{Use Case: Lattice Design}

Generally speaking, synchrotron lattices are constructed from repetitive
blocks comprising of unit cells \& matching sections/dispersion suppressors
for straight sections. The unit cell can be e.g. a FODO cell for beam
transport for colliders or a low emittance cell for synchrotron light
sources. Whereas the matching cells provide for interaction regions
or space for insertion devices.

Hence, from the two outlined Use Cases\c{ } sections \ref{subsec:Matching}
\& \ref{subsec:Periodic_Cell}, two \textquotedbl building blocks\textquotedbl{}
are obtained from which a prototype ring structure with an arbitrary
number of unit cells \& straight sections \textendash{} with adjacent
matching cells \textendash{} can be constructed. After which more
detailed design of the local linear optics can be pursued.

\section{Conclusions}

For a systematic, first principles approach, the formula relevant
for Linear Optics design of Synchrotrons have been derived by Hamiltonian
Dynamics. Equipped with these, the relevant ``Use Cases'' have then
been captured providing a streamlined approach. In particular, to
enable professionals, i.e., Software Engineers, to efficiently prototype
\& architect a CAD Tool for ditto.

We conjecture that the resulting Tools \& Approach will generate a
Paradigm Shift for Robust Linear Optics Design; i.e., better designs
will be found by enabling the Linear Optics Designer to systematically
\& effectively explore the Full Parameter Space for a Prototype Lattice
Design interactively. Besides, the approach might lead to a closer
\& more streamlined approach \& collaboration with Engineers; since
a state-of-the-art design is a matter of Engineering-Science; e.g.
MAX-IV \cite{MAX_IV_saga,MAX_IV_DDR,MAX-IV}.

\newpage{}

\appendix

\part*{Appendices}

\section{Relativistic Hamiltonian\label{sec:Rel_H}}

\subsection{Hamiltonian}

The \textit{relativistic Hamiltonian} for a charged particle with
charge $q$ and energy $E$ in an external electromagnetic field with
vector potential $\bar{A}$ \textendash{} for the co-moving system,
customarily used to model particle accelerators \textendash{} is ($\bar{x}=\left[x,p_{x},y,p_{y},t,-\varepsilon\right]$)
\cite{CERN-88-05,SLS_1997}
\begin{equation}
H\left(\bar{x};s\right)=-p_{s}=-\left(1+h\left(s\right)x\right)\left[\frac{q}{p_{0}}A_{s}+\sqrt{\left(\frac{\varepsilon}{c_{0}}-\frac{q\Phi\left(s\right)}{p_{0}c_{0}}\right)^{2}-\frac{1}{\beta_{0}^{2}\gamma_{0}^{2}}-\left(p_{x}-\frac{qA_{x}\left(s\right)}{p_{0}}\right)^{2}-\left(p_{y}-\frac{qA_{y}\left(s\right)}{p_{0}}\right)^{2}}\right]\label{eq:Rel_H}
\end{equation}
where $\left[\beta,\gamma\right]$ are the relativistic factors
\begin{equation}
\gamma\equiv\frac{1}{\sqrt{1-\beta^{2}}}=\frac{E}{m_{0}c_{0}^{2}},\qquad\varepsilon\equiv\frac{E}{p_{0}}
\end{equation}
and
\begin{equation}
h\left(s\right)\equiv\frac{1}{\rho\left(s\right)}
\end{equation}
is the \textsl{local curvature} for the reference trajectory, see
Fig. \eqref{fig:Co-Moving_Frame}.

Introducing longitudinal coordinates relative to the reference particle
and scaling with $c_{0}$
\begin{equation}
c_{0}T\equiv\left(t-\frac{s}{v_{0}}\right)c_{0},\qquad P_{t}\equiv\frac{E-E_{0}}{p_{0}c_{0}}=\frac{\Delta E}{p_{0}c_{0}}=\frac{\varepsilon-\varepsilon_{0}}{c_{0}}=\frac{\varepsilon}{c_{0}}-\frac{1}{\beta_{0}}
\end{equation}
by the canonical transformation
\begin{equation}
F_{2}\left(t,P_{t}\right)=\left(t-\frac{s}{v_{0}}\right)c_{0}\left(\frac{1}{\beta_{0}}+P_{t}\right)
\end{equation}
gives
\begin{equation}
\begin{cases}
c_{0}T & =\partial_{P_{t}}F_{2}=\left(t-\frac{s}{v_{0}}\right)c_{0}\\
-\varepsilon & =\partial_{t}F_{2}=c_{0}\left(\frac{1}{\beta_{0}}+P_{t}\right)\\
K & =\partial_{s}F_{2}=\frac{1}{\beta_{0}^{2}}+\frac{P_{t}}{\beta_{0}}
\end{cases}
\end{equation}
and the new Hamiltonian is ($\bar{x}=\left[x,p_{x},y,p_{y},c_{0}t,-p_{t}\right]$)
\begin{equation}
H\left(\bar{x};s\right)=\frac{p_{t}}{\beta_{0}}-\left(1+h\left(s\right)x\right)\left[\frac{q}{p_{0}}A_{s}+\sqrt{\left(\frac{1}{\beta_{0}}+p_{t}-\frac{q\Phi\left(s\right)}{p_{0}c_{0}}\right)^{2}-\frac{1}{\beta_{0}^{2}\gamma_{0}^{2}}-\left(p_{x}-\frac{qA_{x}\left(s\right)}{p_{0}}\right)^{2}-\left(p_{y}-\frac{qA_{y}\left(s\right)}{p_{0}}\right)^{2}}\right]\label{eq:Rel_Hamil-1}
\end{equation}
ignoring the constant term $\nicefrac{1}{\beta_{0}^{2}},$using $H$
for the new Hamiltonian, and small letters for the new phase-space
coordinates.

The equations of motions are (\textit{Hamilton's equations}, $\bar{x}\equiv\left[q^{1},p_{1},q^{2},p_{2},\ldots\right]$)
\begin{equation}
\bar{x}'=\left\{ -H,\bar{x}\right\} =-\left(\left(\nabla_{\bar{x}}H\right)\Omega\right)^{\mathrm{T}}=\left[\partial_{p_{1}}H,-\partial_{q^{1}}H,\partial_{p_{2}}H,-\partial_{q^{2}}H,\ldots\right]^{\mathrm{T}}
\end{equation}
where (\textit{Poisson Bracket})
\begin{equation}
\left\{ -H,\cdot\right\} \equiv-\sum_{k}\left[\left(\partial_{q^{k}}H\right)\partial_{p_{k}}-\left(\partial_{p_{k}}H\right)\partial_{q^{k}}\right]
\end{equation}
and (\emph{symplectic form}, from Greek \textit{intertwined})
\begin{equation}
\Omega\equiv\left[\begin{array}{ccc}
\Omega_{x} & 0 & 0\\
0 & \Omega_{y} & 0\\
0 & 0 & \Omega_{s}
\end{array}\right],\qquad\Omega_{x,y,s}=\left[\begin{array}{cc}
0 & 1\\
-1 & 0
\end{array}\right].\label{eq:sympl_form}
\end{equation}

N.B. The upper \& lower indeces are a reflection of the corresponding
differential forms/geometry; i.e., a point of the \textit{cotangent
bundle} $T^{\ast}V$ (i.e., \textit{phase-space}) is specified by
the coordinates $\left[q^{i},p_{j}\right]$.

The \textit{fundamental Poissan brackets} are (Lie algebra)
\begin{equation}
\left\{ q^{i},p_{j}\right\} =\delta_{j}^{i},\qquad\left\{ q^{i},q^{j}\right\} =\left\{ p_{i},p_{j}\right\} =0
\end{equation}
and the Hamiltonian flow's divergence is (vs. a general vector flow)
\begin{equation}
\nabla_{\bar{x}}\cdot\bar{x}'=\nabla_{\bar{x}}\cdot\left\{ -H,\bar{x}\right\} =0
\end{equation}
i.e., the symplectic flow in phase-space is akin to an incompressible
fluid; for a time independent Hamiltonian.

Rather than integrating, etc., it is more expedient and transparent
to utilize the Lie Series solution \cite{Grobner_1,Lie} for the \textit{Poincaré
map} (introduced to celestial mechanics 1960)

\begin{equation}
\bar{x}_{1}=\mathscr{M}\bar{x}_{0}=e^{\mathscr{D}\left(-H\right)}\bar{x}_{0}\equiv\sum_{k=0}^{\infty}\frac{\mathscr{D}^{k}\left(-H\right)}{k!}\bar{x}_{0}
\end{equation}
where (\textit{Lie derivative})
\begin{equation}
\mathscr{D}\left(-H\right)=\left\{ -H,\cdot\right\} .
\end{equation}

\subsection{Multipole Expansion}

The \textit{magnetic multipole expansion} is introduced by (polar
coordinates for $\left[x,y\right]$)
\begin{align}
B_{y}\left(s\right)+iB_{x}\left(s\right) & \equiv\left(B\rho\right)\sum_{n=1}^{\infty}\left(ia_{n}\left(s\right)+b_{n}\left(s\right)\right)\left(re^{i\varphi}\right)^{n-1}\nonumber \\
 & =\left(B\rho\right)\sum_{n=1}^{\infty}\left(ia_{n}\left(s\right)+b_{n}\left(s\right)\right)\left(x+iy\right)^{n-1}
\end{align}
where (\textit{magnetic rigidity}) 
\begin{equation}
\left(B\rho\right)\equiv\frac{p}{q}
\end{equation}
see Fig. \eqref{fig:Magn_Rig}.

The corresponding \textit{vector potential} is obtained from (Poincaré
gauge, $\bar{r}\cdot\bar{A}=0$) \cite{Jackson}
\begin{equation}
\bar{A}\left(\bar{r},t\right)=-\bar{r}\times\intop_{0}^{1}\bar{B}\left(u\bar{r},t\right)udu,\qquad\phi=-\bar{r}\cdot\intop_{0}^{1}\bar{B}\left(u\bar{r},t\right)udu
\end{equation}
which gives (Cartesian coordinates for $\left[x,s\right]$, $h\left(s\right)=0$
)
\begin{equation}
\begin{cases}
A_{x}=A_{y} & =0\\
\frac{q}{p_{0}}A_{s}\left(s\right) & =\mathrm{Re}\sum_{n=1}^{\infty}\frac{1}{n}\left(ia_{n}\left(s\right)+b_{n}\left(s\right)\right)\left(re^{i\varphi}\right)^{n}\\
 & =\mathrm{Re}\sum_{n=1}^{\infty}\frac{1}{n}\left(ia_{n}\left(s\right)+b_{n}\left(s\right)\right)\left(x+iy\right)^{n}.
\end{cases}\label{eq:Mag_Mult_Exp}
\end{equation}

For a \textit{sector bend} (polar coordinates for $\left[x,s\right]$,
arc reference trajectory)
\begin{equation}
\frac{q}{p_{0}}A_{s}=-b_{0}\left(x-\frac{hx^{2}}{2\left(1+hx\right)}\right)=-\frac{b_{0}x\left(2+hx\right)}{2\left(1+hx\right)},\qquad h=b_{0}.
\end{equation}
For completeness, correspondingly, the field, curl$\bar{B}\equiv\bar{\nabla}\times\bar{A}$,
in the curvilinear co-moving frame is \cite{CERN-88-05}

\begin{equation}
B_{x}=\frac{\partial_{z}A_{y}-\partial_{y}A_{z}}{1+h\left(s\right)x},\qquad B_{y}=\frac{h\left(s\right)A_{s}}{1+h\left(s\right)x}+\partial_{x}A_{s}-\frac{\partial_{s}A_{x}}{1+h\left(s\right)x},\qquad B_{z}=\partial_{y}A_{x}-\partial_{x}A_{y}.
\end{equation}

\subsection{Paraxial Approximation}

For the \textit{paraxial approximation} $p_{x,y}\ll1$ the Hamiltonian
simplifies to (not expanded in $p_{t}$ by first factoring out $1+\delta\left(p_{t}\right)$)
\begin{equation}
H\left(\bar{x};s\right)=\frac{\left(p_{x}-\frac{qA_{x}\left(s\right)}{p_{0}}\right)^{2}+\left(p_{y}-\frac{qA_{y}\left(s\right)}{p_{0}}\right)^{2}}{2\left(1+\delta\left(p_{t}\right)\right)}-\left(1+h\left(s\right)x\right)\left(\frac{qA_{s}\left(s\right)}{p_{0}}+1+\delta\left(p_{t}\right)\right)+\frac{q\Phi\left(s\right)}{p_{0}c_{0}}+\ldots\label{eq:H}
\end{equation}
where
\begin{equation}
1+\delta\left(p_{t}\right)=\frac{p\left(p_{t}\right)}{p_{0}}=\sqrt{1+\frac{2p_{t}}{\beta_{0}}+p_{t}^{2}}=\frac{\beta\gamma}{\beta_{0}\gamma_{0}},\qquad\delta\left(p_{t}\right)\equiv\frac{p\left(p_{t}\right)-p_{0}}{p_{0}}
\end{equation}
 By inserting to\textit{ multipole expansion} Eq. (\ref{eq:Mag_Mult_Exp}),
to leading order one obtains (\textit{quadratic Hamiltonian})
\begin{equation}
H_{2}\left(\bar{x};s\right)=\frac{p_{x}^{2}+p_{y}^{2}}{2\left(1+\delta\left(p_{t}\right)\right)}+\frac{K_{x}\left(s\right)x^{2}}{2}-\frac{K_{y}\left(s\right)y^{2}}{2}-\frac{h\left(s\right)xp_{t}}{\beta_{0}}+\frac{p_{t}^{2}}{2\beta_{0}^{2}\gamma_{0}^{2}}+\frac{q\Phi\left(s\right)}{p_{0}c_{0}}\label{eq:Quadr_Hamil}
\end{equation}
where
\begin{equation}
K_{x}\left(s\right)\equiv b_{2}\left(s\right)+h^{2}\left(s\right),\qquad K_{y}\left(s\right)\equiv b_{2}\left(s\right).
\end{equation}

\subsection{Dispersion}

The inhomogeneous term $\nicefrac{h\left(s\right)xp_{t}}{\beta_{0}}$
is removed from the \textit{quadratic Hamiltonian} Eq. (\ref{eq:Quadr_Hamil})
by introducing the \textit{dispersion function} 
\begin{equation}
\bar{x}\left(s,\delta\right)\equiv\eta{}_{x}\left(s\right)\delta\left(p_{t}\right)
\end{equation}
which by inserting into Hamilton's equations
\begin{equation}
\left\{ \begin{aligned}x' & =\partial_{p_{x}}H_{2}=\frac{p_{x}}{1+\delta\left(p_{t}\right)}\\
p_{x}' & =-\partial_{x}H_{2}=-K_{x}\left(s\right)x+h\left(s\right)\delta\left(p_{t}\right)
\end{aligned}
\right.
\end{equation}
gives
\begin{equation}
\eta''{}_{x}+\frac{K_{x}\left(s\right)}{1+\delta\left(p_{t}\right)}\eta{}_{x}=\frac{h\left(s\right)}{1+\delta\left(p_{t}\right)}\label{eq:Disp}
\end{equation}
and the periodic solution can be obtained by solving perturbatively
\begin{equation}
\eta_{x}\left(s\right)=\eta_{x}^{\left(1\right)}\left(s\right)+\eta_{x}^{\left(2\right)}\left(s\right)\delta\left(p_{t}\right)+\ldots
\end{equation}
The \textit{Poincaré map} is ($\bar{x}_{k}\equiv\left[x,p_{x}\right]$)
\begin{equation}
\bar{x}_{k+1}=M\bar{x}_{k}+Dp_{t}
\end{equation}
where
\begin{equation}
M=\left[\begin{array}{cc}
\cos\left(2\pi\nu_{x}\right)+\alpha_{x}\sin\left(2\pi\nu_{x}\right) & \beta_{x}\sin\left(2\pi\nu_{x}\right)\\
-\gamma_{x}\sin\left(2\pi\nu_{x}\right) & \cos\left(2\pi\nu_{x}\right)-\alpha_{x}\sin\left(2\pi\nu_{x}\right)
\end{array}\right],\qquad D=\left[\begin{array}{c}
m_{16}\\
m_{26}
\end{array}\right]
\end{equation}
and the \textit{periodic solution} is given by 
\begin{equation}
\bar{\eta}=M\bar{\eta}+D\Rightarrow\bar{\eta}=\left(I-M\right)^{-1}D
\end{equation}
which gives
\begin{equation}
\bar{\eta}=\left[\begin{array}{cc}
\frac{1+\alpha_{x}\cot\left(\pi\nu_{x}\right)}{2} & \frac{\beta_{x}\cot\left(\pi\nu_{x}\right)}{2}\\
-\frac{\gamma_{x}\cot\left(\pi\nu_{x}\right)}{2} & \frac{1-\alpha_{x}\cot\left(\pi\nu_{x}\right)}{2}
\end{array}\right]\left[\begin{array}{c}
m_{16}\\
m_{26}
\end{array}\right].
\end{equation}
The new Hamiltonian is obtained by the \textit{generating function}
\cite{CERN-88-05}
\begin{equation}
F_{2}\left(x,t,P_{x},P_{t};s\right)=\left(x-\eta_{x}\left(s\right)\delta\left(P_{t}\right)\right)\left(P_{x}+\eta'_{x}\left(s\right)\delta\left(P_{t}\right)\left(1+\delta\left(P_{t}\right)\right)\right)+\frac{1}{2}\eta_{x}\left(s\right)\eta'_{x}\left(s\right)\delta^{2}\left(P_{t}\right)\left(1+\delta\left(P_{t}\right)\right)+c_{0}tP_{t}
\end{equation}
which gives
\begin{equation}
\left\{ \begin{aligned}X & =\partial_{P_{x}}F_{2}=x-\eta{}_{x}\left(s\right)\delta\left(P_{t}\right)\\
p_{x} & =\partial_{x}F_{2}=P_{x}+\eta'_{x}\left(s\right)\delta\left(P_{t}\right)\left(1+\delta\left(P_{t}\right)\right)\\
c_{0}T & =\partial_{P_{t}}F_{2}=c_{0}t-\frac{\frac{1}{\beta_{0}}+p_{t}}{\sqrt{1+\frac{2p_{t}}{\beta_{0}}+p_{t}^{2}}}[\eta_{x}\left(s\right)P_{x}-\eta'_{x}\left(s\right)\left(X+2X\delta\left(P_{t}\right)+\frac{1}{2}\eta_{x}\left(s\right)\delta^{2}\left(P_{t}\right)\right)]\\
p_{t} & =\partial_{c_{0}t}F_{2}=\delta\left(P_{t}\right)\\
K_{2}\left(X,P_{x};s\right) & =H_{2}\left(\bar{x};s\right)+\partial_{s}F_{2}\left(x,P_{x};s\right)
\end{aligned}
\right.
\end{equation}
and 
\begin{equation}
H_{2}\left(\bar{x};s\right)=\frac{p_{x}^{2}+p_{y}^{2}}{2\left(1+p_{t}\right)}+\frac{K_{x}\left(s\right)x^{2}}{2}-\frac{K_{y}\left(s\right)y^{2}}{2}-\frac{\left(h\left(s\right)\eta_{x}\left(s\right)-\frac{1}{\beta_{0}^{2}\gamma_{0}^{2}}\right)p_{t}^{2}}{2}+\frac{q\Phi\left(s\right)}{p_{0}c_{0}}
\end{equation}
using $H$ for the new Hamiltonian and small letters for the new phase-space
coordinates.

N.B.: In the longitudinal plane, dispersion generates amplitude dependence
of the path length \textendash{} due to coupling between the planes
\textendash{} which can e.g. drive synchro-betatron resonances.

\subsection{Momentum Compaction}

Averaging over one turn in the longitudinal plane \textendash{} since
the synchtrotron frequency is much lower than the betatron, $\nu_{s}\ll\nu_{x,y}$
\textendash{} gives (\textit{adiabatic approximation}) \cite{Suzuki}
\begin{align}
\left\langle H_{2}\left(\bar{x};s\right)\right\rangle _{s}= & \frac{p_{x}^{2}+p_{y}^{2}}{2\left(1+\delta\left(p_{t}\right)\right)}+\frac{K_{x}\left(s\right)x^{2}}{2}-\frac{K_{y}\left(s\right)y^{2}}{2}-\frac{\left(\left\langle \frac{h\left(s\right)\eta_{x}\left(s\right)}{2}\right\rangle _{s}-\frac{1}{\beta_{0}^{2}\gamma_{0}^{2}}\right)p_{t}^{2}}{2}+\frac{q\Phi\left(s\right)}{p_{0}c_{0}}+\ldots\nonumber \\
= & \frac{p_{x}^{2}+p_{y}^{2}}{2\left(1+\delta\left(p_{t}\right)\right)}+\frac{K_{x}\left(s\right)x^{2}}{2}-\frac{K_{y}\left(s\right)y^{2}}{2}-\frac{\eta_{\mathrm{c}}\left(p_{t}\right)p_{t}^{2}}{2}+\frac{q\Phi\left(s\right)}{p_{0}c_{0}}+\ldots\label{eq:H_6D}
\end{align}
where $\eta_{\mathrm{c}}\left(p_{t}\right)$ (\textit{closed orbit
phase slip factor}\textit{\emph{)}}
\begin{equation}
\eta_{\mathrm{c}}\left(p_{t}\right)\equiv\frac{\nicefrac{\left(T-T_{0}\right)}{T_{0}}}{\nicefrac{\left(E-E_{0}\right)}{E_{0}}}=\frac{\nicefrac{\left(C-C_{0}\right)}{T_{0}}}{\nicefrac{\left(E-E_{0}\right)}{E_{0}}}-\frac{\beta-\beta_{0}}{\beta_{0}}=\alpha_{\mathrm{c}}\left(p_{t}\right)-\frac{1}{\beta_{0}^{2}\gamma_{0}^{2}}
\end{equation}
and (\textit{momentum compaction})
\begin{equation}
\alpha_{\mathrm{c}}\left(p_{t}\right)\equiv\frac{\nicefrac{\left(C-C_{0}\right)}{C_{0}}}{\nicefrac{\left(E-E_{0}\right)}{E_{0}}}=\frac{1}{C}\ointop\frac{\eta_{x}\left(s\right)}{\rho\left(s\right)}ds=\alpha_{\mathrm{c}}^{\left(1\right)}+\alpha_{\mathrm{c}}^{\left(2\right)}\frac{E-E_{0}}{E_{0}}+\ldots\label{eq:alpha_c}
\end{equation}
The equations of motion for the time-of-flight to leading order for
$p_{x}=p_{y}=0$ is
\begin{equation}
c_{0}t'=\frac{c_{0}}{\dot{s}}=\partial_{-p_{t}}H_{2}=\eta_{\mathrm{c}}p_{t}=\eta_{\mathrm{c}}\frac{\Delta E}{p_{0}c_{0}}=\eta_{\mathrm{c}}\frac{\Delta E}{\beta_{0}E_{0}}
\end{equation}
which gives
\begin{equation}
\beta_{0}c_{0}t'=\eta_{\mathrm{c}}\frac{\Delta E}{E_{0}}.
\end{equation}

\subsection{Betatron Motion: Action-Angle Coordinates\label{subsec:Action-Angle_Coord}}

The Hamiltonian for the horizontal plane is Eq. \eqref{eq:H_2} 
\begin{equation}
H_{2}\left(\bar{x};s\right)=\frac{p_{x}^{2}}{2\left(1+p_{t}\right)}+\frac{K\left(s\right)x^{2}}{2}
\end{equation}
and the \textit{equations of motion} are (Hamilton's equations)
\begin{equation}
\left\{ \begin{aligned}x' & =\partial_{p_{x}}H_{2}=\frac{p_{x}}{1+p_{t}}\\
p'_{x} & =-\partial_{x}H_{2}=-K\left(s\right)x
\end{aligned}
\right.
\end{equation}
which can be combined into (\textit{Hill's equation})
\begin{equation}
x''+\frac{K\left(s\right)}{1+p_{t}}x=0.\label{eq:Hill's_Eq.}
\end{equation}
The \textit{pseudo-harmonic oscillator} ansatz (\footnote{Similar to the WKB (Wentzel\textendash Kramers\textendash Brillouin)
approximation in Quantum Mechanics; aka Liouville\textendash Green
method \cite{WKB}.})
\begin{equation}
x\left(s;\delta\right)=\sqrt{\beta_{x}\left(s;\delta\right)}e^{\pm i\psi_{x}\left(s;\delta\right)},\qquad\alpha_{x}\left(s;\delta\right)\equiv-\frac{1}{2}\partial_{s}\beta_{x}\left(s;\delta\right)
\end{equation}
gives
\begin{equation}
\left(\sqrt{\beta_{x}}\right)''+K\left(s\right)\sqrt{\beta_{x}}-\sqrt{\beta_{x}}\left(\psi'_{x}\right)^{2}\pm i\left(\sqrt{\beta_{x}}\psi''_{x}+\frac{\beta'_{x}}{\sqrt{\beta_{x}}}\psi'_{x}\right)=0
\end{equation}
which yields the relations
\begin{equation}
\left\{ \begin{aligned}\left(\sqrt{\beta_{x}}\right)''+K\left(s\right)\sqrt{\beta_{x}}-\sqrt{\beta_{x}}\left(\psi'_{x}\right)^{2} & =0\\
\frac{\psi''_{x}}{\psi'_{x}} & =-\frac{\beta'_{x}}{\beta_{x}}
\end{aligned}
\right.\label{eq:beta_function}
\end{equation}
and integrating the second equation gives
\begin{equation}
\psi{}_{x}\left(s;\delta\right)=\mu_{x}\left(s;\delta\right)+\phi_{x},\qquad\mu_{x}\left(s;\delta\right)\equiv\int_{0}^{s}\frac{du}{\beta_{x}\left(u;\delta\right)}\label{eq:phase_advance}
\end{equation}
where $\mu_{x}\left(s\right)$ is the Courant \& Snyder \textit{phase-advance}.
The global parameters $\left[\alpha,\beta,\mu\right]$ are called
the \textit{Twiss parameters} for the lattice \cite{C-S,Twiss_Frank}.

\textit{Action-angle coordinates} $\left[J_{x},\psi_{x}\right]$ are
introduced by the generating function \cite{Action_Angle}
\begin{equation}
F_{1}\left(x,\psi_{x};s\right)=-\frac{x^{2}}{2\beta_{x}\left(s;p_{t}\right)}\left(\tan\left(\psi_{x}\right)+\alpha_{x}\left(s;p_{t}\right)\right)
\end{equation}
with

\begin{equation}
\left\{ \begin{aligned}p_{x} & =-\partial_{x}F_{1}=-\frac{x}{\beta_{x}}\left(\tan\left(\psi_{x}\right)+\alpha_{x}\right)\\
J_{x} & =-\partial_{\psi_{x}}F_{1}=\frac{x^{2}}{2\beta_{x}\cos^{2}\left(\psi_{x}\right)}\\
K_{2} & =H_{2}+\partial_{s}F_{1}
\end{aligned}
\right.
\end{equation}
Inverting the first and second equation gives
\begin{equation}
\left\{ \begin{aligned}\psi_{x}\left(x,p_{x}\right) & =\mu_{x}\left(s;\delta\right)-\atan2\left(\frac{\tilde{p}_{x}}{\tilde{x}}\right)=-\atan2\left(\frac{\alpha_{x}\left(s;\delta\right)x+\beta_{x}\left(s;\delta\right)p_{x}}{x}\right)\\
2J_{x}\left(x,p_{x}\right) & =\left|\tilde{x}\right|^{2}=\tilde{x}^{2}+\tilde{p}_{x}^{2}=\bar{x}^{\mathrm{T}}\left(AA^{\mathrm{T}}\right)^{-1}\left(s\right)\bar{x}=\gamma_{x}\left(s;\delta\right)x^{2}+2\alpha_{x}\left(s;\delta\right)xp_{x}+\beta_{x}\left(s;\delta\right)p_{x}^{2}
\end{aligned}
\right.\label{eq:action_angle_var_2}
\end{equation}
where (\textit{Floquet space})
\begin{equation}
\tilde{x}=A^{-1}\bar{x}
\end{equation}
with 
\begin{equation}
A=\left[\begin{array}{cc}
\sqrt{\beta_{x}} & 0\\
-\frac{\alpha_{x}}{\sqrt{\beta_{x}}} & \frac{1}{\sqrt{\beta_{x}}}
\end{array}\right],\qquad A^{-1}=\left[\begin{array}{cc}
\frac{1}{\sqrt{\beta_{x}}} & 0\\
\frac{\alpha_{x}}{\sqrt{\beta_{x}}} & \sqrt{\beta_{x}}
\end{array}\right]\label{eq:C_S_phase_advance}
\end{equation}
see Fig. \eqref{fig:Phase-Space_Fl-Space}.

Conversely
\begin{equation}
\left\{ \begin{aligned}x\left(s\right) & =\sqrt{2J_{x}\beta_{x}\left(s;\delta\right)}\cos\left(\psi_{x}\left(s;p_{t}\right)\right)\\
p_{x}\left(s\right) & =-\sqrt{\frac{2J_{x}}{\beta_{x}\left(s;\delta\right)}}\left(\sin\left(\psi_{x}\left(s;p_{t}\right)\right)+\alpha_{x}\left(s;p_{t}\right)\cos\left(\psi_{x}\left(s;p_{t}\right)\right)\right)\\
 & =-\sqrt{2J_{x}\gamma_{x}\left(s;\delta\right)}\sin\left(\psi_{x}\left(s;p_{t}\right)+\arctan\left(\alpha_{x}\left(s;p_{t}\right)\right)\right)
\end{aligned}
\right.\label{eq:betatron_motion_2}
\end{equation}
see Fig. \eqref{fig:Phase-Space_Ellips}.

The \emph{action} in \emph{diagonal form} is
\begin{equation}
2J_{x}=\tilde{x}^{\mathrm{T}}\tilde{x}=\lambda_{+}\tilde{x}^{2}+\lambda_{-}\tilde{p}_{x}^{2}=\frac{\tilde{x}^{2}}{a^{2}}+\frac{\tilde{p}_{x}^{2}}{b^{2}},\qquad\lambda_{\pm}=h\pm\sqrt{h^{2}-1},\qquad h=\frac{\beta_{x}+\gamma_{x}}{2}
\end{equation}
i.e., the \emph{ellipse parameters} for the \emph{quadratic form}
are
\begin{align}
a,b & =\sqrt{\lambda_{\pm}}=\sqrt{h\pm\sqrt{h^{2}-1}}=\frac{1}{\sqrt{2}}\left(\sqrt{h+1}\pm\sqrt{h-1}\right),\nonumber \\
\tan\left(2\xi_{x}\right) & =\frac{2a_{12}}{a_{11}-a_{22}}=\frac{2\alpha_{x}}{\gamma_{x}-\beta_{x}},\nonumber \\
\mathrm{Area} & =\frac{\pi ab}{4}=\frac{\pi}{4}
\end{align}
where $\psi_{x}$ is the major's angle relative to the $x$-axis;
a phase-space rotation which diagonalizes $AA^{\mathrm{T}}$. Note,
$\xi_{x}$ is not the phase advance $\psi_{x}$; which is determined
by the particular choice of $A$, Eq. \eqref{eq:C-S_Phase-Advance}.

The equation for the ellipse with the transformed coordinates is

\begin{equation}
\left\{ \begin{aligned}\tilde{x}\left(s\right) & =a\cos\left(\tilde{\psi}_{x}\right)\\
\tilde{p}_{x}\left(s\right) & =-b\sin\left(\tilde{\psi_{x}}\right)
\end{aligned}
\right..
\end{equation}

The \emph{Poincaré map} is
\begin{equation}
\bar{x}_{k+1}=M\bar{x}_{k}
\end{equation}
where
\begin{equation}
M=\left[\begin{array}{cc}
\cos\left(\mu_{x}\right)+\alpha_{x}\sin\left(\mu_{x}\right) & \beta_{x}\sin\left(\mu_{x}\right)\\
-\gamma_{x}\sin\left(\mu_{x}\right) & \cos\left(\mu_{x}\right)-\alpha_{x}\sin\left(\mu_{x}\right)
\end{array}\right]
\end{equation}
since (diagonal form)
\begin{equation}
M=AR\left(\mu_{x}\right)A^{-1},\qquad R\left(\mu_{x}\right)\equiv\left[\begin{array}{cc}
\cos\left(\mu_{x}\right) & \sin\left(\mu_{x}\right)\\
-\sin\left(\mu_{x}\right) & \cos\left(\mu_{x}\right)
\end{array}\right].
\end{equation}
The new Hamiltonian is
\begin{equation}
K_{2}\left(J_{x},\psi_{x};s\right)=H_{2}+\frac{\alpha_{x}xp_{x}}{\beta_{x}}-\frac{\alpha'_{x}x^{2}}{2\beta_{x}}=\frac{J_{x}}{\beta_{x}\left(s;p_{t}\right)}
\end{equation}
where Eq. \eqref{eq:Hill's_Eq.} has been used to simplify the expression.

The equations of motion are (Hamilton's equations)
\begin{equation}
\left\{ \begin{aligned}J'_{x} & =-\partial_{\phi_{x}}K_{2}=0\\
\psi'_{x} & =\partial_{J_{x}}K_{2}=\frac{1}{\beta_{x}\left(s;p_{t}\right)}
\end{aligned}
\right.
\end{equation}
and by integrating
\begin{equation}
\left\{ \begin{aligned}J_{x} & =\mathrm{cst}.\\
\psi{}_{x} & =\int_{0}^{s}\frac{du}{\beta_{x}\left(u;p_{t}\right)}+\phi_{x}=\mu_{x}\left(s;\delta\right)+\phi_{x}
\end{aligned}
\right.
\end{equation}
where $\left[\phi_{x},J_{x}\right]$ are the \textit{constants of
motion}.

Averaging the Hamiltonian over one turn
\begin{equation}
\left\langle K_{2}\left(J_{x},\psi_{x};s\right)\right\rangle =\frac{J_{x}}{C}\int_{0}^{C}\frac{ds}{\beta_{x}\left(s;p_{t}\right)}=\frac{J_{x}}{\lambda_{x}\left(p_{t}\right)}=\frac{2\pi\left(\nu_{x}+\xi_{x}p_{t}\right)J_{x}}{C}\label{eq:avg_Hamiltonian}
\end{equation}
where
\begin{equation}
\frac{1}{\lambda_{x}\left(\delta\right)}\equiv\left\langle \frac{1}{\beta_{x}\left(s;p_{t}\right)}\right\rangle =\frac{2\pi\nu_{x}\left(\delta\right)}{C}=\frac{2\pi\left(\nu_{x}+\xi_{x}p_{t}\right)}{C}\label{eq:betatron_tune}
\end{equation}

\begin{elabeling}{00.00.0000}
\item [{and}]~
\begin{elabeling}{00.00.0000}
\item [{$C$}] circumference,
\item [{$2\pi\nu_{x}\equiv\mu_{x}$}] horizontal tune,
\item [{$\xi_{x}$}] chromaticity.
\end{elabeling}
\end{elabeling}
The \textit{chromaticity} generates momentum dependence of the betatron
tune; i.e., coupling between the planes. Conversely, it also generates
a quadratic amplitude dependence of the path length

\begin{equation}
\Delta s'=-\partial_{\delta}\left\langle K_{2}\left(J_{x},\psi_{x};s\right)\right\rangle =-\frac{2\pi\xi_{x}J_{x}}{C}=-\frac{\pi\xi_{x}\hat{x}^{2}}{C\beta_{x}},\label{eq:path_length_dep_3}
\end{equation}
see section \ref{subsec:Path-Length-Amplitude}.

The averaged equations of motion are
\begin{equation}
\left\{ \begin{aligned}J'_{x} & =-\partial_{\phi_{x}}\left\langle K_{2}\left(J_{x},\phi_{x};s\right)\right\rangle =0\\
\phi'_{x} & =\partial_{J_{x}}\left\langle K_{2}\left(J_{x},\phi_{x};s\right)\right\rangle =\frac{1}{\lambda_{x}\left(\delta\right)}=\frac{2\pi\left(\nu_{x}+\xi_{x}p_{t}\right)}{C}
\end{aligned}
\right.
\end{equation}
and the corresponding action is ($\Rightarrow\alpha_{x}=0$) 
\begin{equation}
J_{x}=\frac{1}{2}\left(\frac{x^{2}}{\lambda_{x}\left(\delta\right)}+\lambda_{x}\left(\delta\right)p_{x}^{2}\right).
\end{equation}
The averaged Hamiltonian in the original phase space coordinates is
\begin{equation}
\left\langle K_{2}\left(x,p_{x};s\right)\right\rangle =\frac{J_{x}}{\lambda_{x}\left(\delta\right)}=\frac{p_{x}^{2}}{2}+\frac{x^{2}}{2\lambda_{x}^{2}\left(\delta\right)}
\end{equation}
and the generalized Hill's equation \eqref{eq:Hill's_Eq.}
\begin{equation}
x''+\frac{1}{\lambda_{x}^{2}\left(p_{t}\right)}x=0\label{eq:gen_Hill's_eq}
\end{equation}
has the solution
\begin{equation}
\left\{ \begin{aligned}x\left(s\right) & =\sqrt{2J_{x}}\cos\left(\frac{s}{\lambda_{x}\left(\delta\right)}+\phi_{x}\right)\\
p_{x}\left(s\right) & =-\frac{\sqrt{2J_{x}}}{\lambda_{x}\left(p_{t}\right)}\sin\left(\frac{s}{\lambda_{x}\left(\delta\right)}+\phi_{x}\right)
\end{aligned}
\right..
\end{equation}

Similar expressions hold for the vertical plane.

\subsection{Dispersion Action\label{subsec:Dispersion_Action}}

Similarly to the action-angle coordinates Eq. \eqref{eq:action_angle_var_2},
for the dispersion $\bar{\eta}\left(s\right)$ (particular solution)
one may introduce
\begin{equation}
\bar{x}=\bar{\eta}\left(s\right)\delta,\qquad\tilde{\eta}=A^{-1}\bar{\eta}=\left[\begin{array}{c}
\sqrt{\mathscr{H}_{x}\left(s\right)}\cos\left(\varphi_{x}\left(s\right)\right)\\
-\sqrt{\mathscr{H}_{x}\left(s\right)}\sin\left(\varphi_{x}\left(s\right)\right)
\end{array}\right]\label{eq:disp_action}
\end{equation}
where the \textit{dispersion action-angle coordinates} $\left[\mathscr{H}_{x},\varphi_{x}\right]$
are
\begin{equation}
\left\{ \begin{aligned}\mathscr{H}_{x}\left(s\right) & \equiv2J_{x}\left(\eta{}_{x},\eta'{}_{x}\right)=\gamma_{x}\left(s;\delta\right)\eta_{x}^{2}+2\alpha_{x}\left(s;\delta\right)\eta{}_{x}\eta'{}_{x}+\beta_{x}\left(s;\delta\right)\left(\eta'{}_{x}\right)^{2}=\left|\tilde{\eta}\right|^{2}=\tilde{\eta}_{x}^{2}+{\tilde{\eta}'}{}_{x}^{2}\\
\varphi_{x}\left(s\right) & \equiv-\atan2\left(\frac{\tilde{\eta}'_{x}}{\tilde{\eta}_{x}}\right)
\end{aligned}
\right.
\end{equation}
with the Hamiltonian
\begin{equation}
K_{2}\left(\mathscr{H}_{x},\varphi_{x};s\right)=\frac{\mathscr{H}_{x}}{\beta_{x}\left(s\right)}-\frac{\sqrt{\mathscr{H}_{x}}\cos\left(\mu_{x}+\varphi_{x}\right)}{\rho\left(s\right)}
\end{equation}
i.e., a \textit{Floquet space rotation} for drifts ($\rho\left(s\right)\rightarrow\infty$);
whereas for dipoles the increase of the magnitude of the dispersion
action $\left|\mathscr{H}_{x}\right|$ is governed by
\begin{equation}
\eta'{}_{x}=\frac{s}{\rho\left(s\right)},\qquad\eta{}_{x}=\int_{0}^{s}\frac{u}{\rho\left(u\right)}du.
\end{equation}
For example, a momentum change $\delta$ for a particle moving along
the design orbit $J_{x}=0,\delta=0$ causes a Closed Orbit shift (by
e.g. an RF cavity, radiation, or Touschek (inelestic) scattering)
\begin{equation}
x\left(s\right)=\eta_{x}\left(s\right)\delta,\qquad p_{x}\left(s\right)=x'\left(s\right)\left(1+p_{t}\right)=\eta'_{x}\left(s\right)\delta+\mathcal{O}\left(\delta^{2}\right)
\end{equation}
which generates a betatron oscillation with the amplitude
\begin{equation}
2J_{x}=\left[\gamma_{x}\eta_{x}^{2}+2\alpha_{x}\eta_{x}\eta'_{x}+\beta_{x}\left(\eta'_{x}\right)^{2}\right]p_{t}^{2}=\mathscr{H}_{x}p_{t}^{2}.
\end{equation}
Hence, the objective of \textit{linear optics design} for synchrotron
light sources is to \textit{control \& minimize $\mathscr{H}_{x}$}
in the quest towards the ultimate limit, aka \textit{diffraction limit}.

\subsubsection{Path Length Amplitude Dependence\label{subsec:Path-Length-Amplitude}}

In the longitudinal plane, the dispersion generates amplitude dependence
of the path length Eq. \eqref{eq:path_length_dep_3}
\begin{equation}
\Delta s\left(s\right)=\eta'{}_{x}\left(s\right)x\left(s\right)-\eta{}_{x}\left(s\right)p_{x}\left(s\right)=\sqrt{2J_{x}\mathscr{H}_{x}\left(s\right)}\sin\left(\psi_{x}\left(s\right)-\varphi_{x}\left(s\right)\right)\label{eq:path_length_dep_4}
\end{equation}
where Eqs. \eqref{eq:disp_action} has been used, which leads to an
orbit $\Delta s_{0}$ (phase shift)
\begin{equation}
\sqrt{2J_{x}\mathscr{H}_{x,0}}\sin\left(\phi_{x}-\varphi_{x,0}\right).
\end{equation}
and an oscillation with amplitude $\sqrt{2J_{x}\mathscr{H}_{x}\left(s\right)}$
driven by the betatron motion. A reflection of the 6D phase-space
dynamics being governed by a symplectic flow.

\section{Radiation Effects\label{sec:Rad_Effects}}

\subsection{Statistical Moments}

For a beam \textendash{} a distribution of particles \textendash{}
the Statistical 2nd Moments are \cite{Emittance}
\begin{equation}
\Sigma\equiv\left\langle \bar{x}\otimes\bar{x}\right\rangle =\left\langle x_{i}x_{j}\right\rangle =\left[\begin{array}{cccc}
\sigma_{x}^{2} & \sigma_{xp_{x}} & \cdots & \sigma_{xp_{t}}\\
\sigma_{xp_{x}} & \sigma_{p_{x}}^{2} & \cdots & \sigma_{p_{x}p_{t}}\\
\vdots & \vdots & \ddots & \vdots\\
\sigma_{xp_{t}} & \sigma_{p_{x}p_{t}} & \cdots & \sigma_{p_{t}}^{2}
\end{array}\right]
\end{equation}
which are propagated by
\begin{equation}
\Sigma\rightarrow M\Sigma M^{\mathrm{T}}.
\end{equation}
The (eigen) emittances $\varepsilon_{k}$ are defined in Floquet space
and are related by
\begin{equation}
\Sigma=\mathrm{Diag}\left[\sqrt{\varepsilon_{x},}\sqrt{\varepsilon_{x}},\sqrt{\varepsilon_{y}},\sqrt{\varepsilon_{y}},\sqrt{\varepsilon_{t}},\sqrt{\varepsilon_{t}}\right]AA^{\mathrm{T}}\mathrm{Diag}\left[\sqrt{\varepsilon_{x},}\sqrt{\varepsilon_{x}},\sqrt{\varepsilon_{y}},\sqrt{\varepsilon_{y}},\sqrt{\varepsilon_{t}},\sqrt{\varepsilon_{t}}\right]
\end{equation}
which for mid-plane symmetry simplifies to
\begin{equation}
\varepsilon_{k}\:\mathrm{\mathrm{[m\cdot rad]}}=\sqrt{\sigma_{q_{k}}^{2}\sigma_{p_{k}}^{2}-\sigma_{q_{k}p_{k}}^{2}}
\end{equation}
since
\begin{equation}
AA^{\mathrm{T}}=\left[\begin{array}{ccc}
A_{x}A_{x}^{\mathrm{T}} & 0 & 0\\
0 & A_{y}A_{y}^{\mathrm{T}} & 0\\
0 & 0 & A_{s}A_{s}^{\mathrm{T}}
\end{array}\right],\qquad A_{k}A_{k}^{\mathrm{T}}=\left[\begin{array}{cc}
\beta_{k} & -\alpha_{k}\\
-\alpha_{k} & \gamma_{k}
\end{array}\right].
\end{equation}

The Particle Distribution is generally approximated by a Gaussian
\begin{equation}
\rho\left(\bar{x}\right)=\frac{1}{\left(2\pi\right)^{3}\Pi{}_{k}\varepsilon_{k}}e^{-\sum_{k}\frac{2J_{k}}{\varepsilon_{k}}}.
\end{equation}

\subsection{Synchrotron Integrals\label{subsec:Synchr-Int}}

The radiation effects - damping \& quantum fluctuations - are governed
by the equations \cite{Sands} (Eqs. (4.3)-(4.4), p. 98, and (5.11)-(5.20),
p. 117-118)
\begin{equation}
\left\{ \begin{aligned}B\rho & =\frac{p}{e}\approx\frac{E}{ec_{0}}\\
P_{\gamma}\:\mathrm{\mathrm{[eV/sec]}} & =\frac{C_{\gamma}c_{0}^{3}e^{2}E^{2}B^{2}}{2\pi}\approx\frac{C_{\gamma}c_{0}E^{4}}{2\pi\rho^{2}}\\
N\:\mathrm{[sec^{-1}\mathrm{]}} & =\int_{0}^{\infty}n\left(u\right)du=\frac{15\sqrt{3}P_{\gamma}}{8u_{\mathrm{c}}}\\
\left\langle u^{2}\right\rangle \:\mathrm{[(eV)^{2}\mathrm{]}} & =\frac{11u_{\mathrm{c}}^{2}}{27}\\
N\left\langle u^{2}\right\rangle \:[(eV)^{2}\mathrm{/sec]} & =\int_{0}^{\infty}u^{2}n\left(u\right)du=C_{\mathrm{u}}\varepsilon_{\mathrm{c}}P_{\gamma}=\frac{3C_{\mathrm{u}}C_{\gamma}\hbar c_{0}^{2}E^{7}}{4\pi\left(m_{\mathrm{e}}c_{0}^{2}\right)^{3}\left|\rho^{3}\right|}
\end{aligned}
\right.
\end{equation}

\begin{elabeling}{00.00.0000}
\item [{where}]~
\begin{elabeling}{00.00.0000}
\item [{$N$}] total quanta emmission rate,
\item [{$n\left(u\right)$}] quanta with energy $u$ emitted per unit time,
\item [{$u_{\mathrm{c}}$}] Critical Photon Energy;
\end{elabeling}
\end{elabeling}
with the constants (Eqs. (4.2), p. 98, (5.3), p. 115, and (5.19),
p. 118 \cite{Sands})
\begin{equation}
C_{\gamma}\:\mathrm{[m\,(eV)^{-3}\mathrm{]}}\equiv\frac{4\pi r_{\mathrm{e}}}{3\left(m_{\mathrm{e}}c_{0}^{2}\right)^{3}}=8.845\times10^{-5},\qquad C_{\mathrm{u}}\:\mathrm{[.]}\equiv\frac{55}{24\sqrt{3}}=1.323,\qquad u_{\mathrm{c}}\:\mathrm{[eV]}\equiv\hbar\omega_{\mathrm{c}}=\frac{3\hbar c_{0}\gamma^{3}}{2\rho}
\end{equation}
and (Classical Electron Radius and Fine Structure Constant)
\begin{equation}
r_{\mathrm{e}}\:\mathrm{[m]}\equiv\frac{e^{2}}{4\pi\varepsilon_{0}m_{\mathrm{e}}c_{0}^{2}}=\frac{\alpha\hbar}{m_{\mathrm{e}}c_{0}}=2.817\times10^{-15},\qquad\alpha\:\mathrm{[.]}\equiv\frac{e^{2}}{4\pi\varepsilon_{0}\hbar c_{0}}=\frac{1}{137.036}
\end{equation}
By introducing (Synchrotron Integrals \cite{Sands,Sands_1973})
\begin{equation}
I_{2}\equiv\ointop\frac{1}{\rho^{2}\left(s\right)}\totd s,\qquad I_{3}\equiv\ointop\frac{1}{\left|\rho^{3}(s)\right|}\totd s,\qquad I_{4}\equiv\ointop\frac{\eta_{x}\left(s\right)}{\rho\left(s\right)}\left(\frac{1}{\rho^{2}\left(s\right)}+2b_{2}\left(s\right)\right)\totd s,\qquad I_{5}\equiv\ointop\frac{\mathscr{H}_{x}(s)}{\left|\rho^{3}(s)\right|}\totd s\label{eq:synchr_int}
\end{equation}
the global radiation properties for a lattice can be obtained from
((4.51)-(4.53), p. 110, (5.36)-(5.46), p. 122-124 ref. \cite{Sands})
\begin{equation}
\left\{ \begin{aligned}U_{0}\:\mathrm{[eV/turn]} & =\ointop\frac{P_{\gamma}}{c_{0}}\totd s=\frac{C_{\gamma}E_{0}^{4}I_{2}}{2\pi}\\
\varepsilon_{x}\:\mathrm{[m}\cdot\mathrm{rad]} & =C_{\mathrm{q}}\gamma^{2}\frac{I_{5}}{j_{x}I_{2}}=C_{\mathrm{q}}\gamma^{2}\frac{I_{5}}{I_{2}-I_{4}}\\
 & =1467.5\times10^{-9}\cdot E^{2}\:\mathrm{[GeV]}\cdot\frac{I_{5}}{I_{2}-I_{4}}\\
\sigma_{p_{t}}^{2}\:\mathrm{[.]} & =C_{\mathrm{q}}\gamma^{2}\frac{I_{3}}{j_{p_{t}}I_{2}}=C_{\mathrm{q}}\gamma^{2}\frac{I_{3}}{2I_{2}+I_{4}}\\
 & =1467.5\times10^{-9}\cdot E^{2}\:\mathrm{[GeV]}\cdot\frac{I_{3}}{2I_{2}+I_{4}}\\
\left[\alpha_{x},\alpha_{y},\alpha_{p_{t}}\right]\:\mathrm{[1/sec]} & =j_{k}\alpha_{0}=j_{k}\frac{\left\langle P_{\gamma}\right\rangle }{2E_{0}}=\frac{U_{0}}{2E_{0}T_{0}}\left[1-\frac{I_{4}}{I_{2}},1,2+\frac{I_{4}}{I_{2}}\right]\\
\left[j_{x},j_{y},j_{p_{t}}\right]\:\mathrm{[.]} & =\left[1-\frac{I_{4}}{I_{2}},1,2+\frac{I_{4}}{I_{2}}\right]\\
\left[\tau_{x},\tau_{y},\tau_{p_{t}}\right]\:\mathrm{[sec]} & =\frac{1}{\alpha_{k}}=\frac{2E_{0}}{j_{k}\left\langle P_{\gamma}\right\rangle }=\frac{2T_{0}E_{0}}{j_{k}U_{0}}=\frac{4\pi T_{0}}{C_{\gamma}E_{0}^{3}j_{k}I_{2}}\\
\left\langle P_{\gamma}\right\rangle \:\mathrm{[eV/sec]} & =\frac{1}{C}\ointop P_{\gamma}\totd s=\frac{C_{\gamma}E_{0}^{4}I_{2}}{2\pi T_{0}}=\frac{U_{0}}{T_{0}}\\
Q_{p_{t}}\:\mathrm{[(eV)^{2}/sec]} & =N\left\langle \left\langle u^{2}\right\rangle \right\rangle _{s}=\frac{1}{C}\ointop N\left\langle u^{2}\right\rangle \totd s=\frac{3C_{\mathrm{u}}\hbar c_{0}\gamma^{3}\left\langle P_{\gamma}\right\rangle I_{3}}{2I_{2}}=\frac{4\sigma_{p_{t}}^{2}}{\tau_{p_{t}}}
\end{aligned}
\right.\label{eq:eps_x_1}
\end{equation}

\begin{elabeling}{00.00.0000}
\item [{where}]~
\begin{elabeling}{00.00.0000}
\item [{$E_{0}$}] reference energy,
\item [{$U_{0}$}] energy loss per turn,
\item [{$\varepsilon_{x}$}] horizontal emittance,
\item [{$\sigma_{p_{t}}$}] momentum spread,
\item [{$\alpha_{k}$}] damping coefficient,
\item [{$j_{k}$}] partition number,
\item [{$\tau_{k}$}] damping time,
\item [{$\gamma$}] relativistic factor,
\item [{$T_{0}=\nicefrac{C}{c_{0}}$}] revolution time,
\item [{$C$}] circumference;
\end{elabeling}
\end{elabeling}
and (Quantum Constant, Eq. (5.46), p. 124 \cite{Sands}, and Reduced
Compton Wavelength)
\begin{equation}
C_{\mathrm{q}}\:\mathrm{[m]}\equiv\frac{55\lambdabar}{32\sqrt{3}},\qquad\lambdabar\equiv\frac{\hbar}{m_{\mathrm{e}}c_{0}}.
\end{equation}
The partition numbers are governed by the sum rule \cite{Robinson}
\begin{equation}
j_{x}+j_{y}+j_{p_{t}}=4.
\end{equation}
In other words, the Synchrotron Integrals provide a convenient parametrization
of the global properties of the lattice; in terms of the Linar Optics
\cite{Sands_1973}.

\subsection{Diffusion Coefficients}

For the general case the Beam Envelope Formalism can be generalized
\cite{Beam_Envelope}. However, a more straightforward \& transparent
Geometric Approach is to model Classical Radiation by generalizing
to a Vector Flow and Quantum Fluctuations by the Diffusion Coefficients;
and then evaluate the effect on the Linear Actions (Invariants) \cite{Chao}.

The emittance $\varepsilon_{k}$ in 6D Floquet space is governed by
the diffusion equation
\begin{equation}
2J'_{k}=-\frac{2T_{0}}{\tau_{k}}2J_{k}+D_{J_{k}},\qquad D_{J_{k}}\equiv\left\langle d_{J_{k}}\left(s\right)\right\rangle 
\end{equation}

\begin{elabeling}{00.00.0000}
\item [{where}] 

\begin{elabeling}{00.00.0000}
\item [{$D_{J_{k}},d_{J_{k}}\left(s\right)$}] global \& local diffusion
coefficients for the action coordinate,
\item [{$\left[\tau_{x},\tau_{y},\tau_{p_{t}}\right]$}] damping times
for the phase space coordinates (hence the factor 2);
\end{elabeling}
\end{elabeling}
which has the stationary solution
\begin{equation}
2J_{k}=\frac{\tau_{k}D_{J_{k}}}{2T_{0}}.
\end{equation}
The beam emittance is the dynamic equilibrium
\begin{equation}
\varepsilon_{k}=\left\langle 2J_{k}\right\rangle =\frac{\tau_{k}D_{J_{k}}}{2T_{0}}=\frac{E_{0}D_{J_{k}}}{j_{k}U_{0}}.
\end{equation}

The Diffusion Matrix in phase-space is
\begin{align}
D\equiv & \mathrm{Diag}\left[\sqrt{D_{J_{x}}},\sqrt{D_{J_{x}}},\ldots,\sqrt{D_{J_{t}}},\sqrt{D_{J_{t}}}\right]AA^{\mathrm{T}}\mathrm{Diag}\left[\sqrt{D_{J_{x}}},\sqrt{D_{J_{x}}},\ldots,\sqrt{D_{J_{t}}},\sqrt{D_{J_{t}}}\right]\nonumber \\
= & \left[\begin{array}{cccc}
D_{xx} & D_{xp_{x}} & \cdots & D_{xp_{t}}\\
D_{xp_{x}} & D_{p_{x}p_{x}} & \cdots & D_{p_{x}p_{t}}\\
\vdots & \vdots & \ddots & \vdots\\
D_{xp_{s}} & D_{p_{x}p_{s}} & \cdots & D_{p_{t}p_{t}}
\end{array}\right]
\end{align}
which is related to the Synchrotron Integrals by
\begin{equation}
\left\{ \begin{aligned}D_{J_{x}}\:\mathrm{\mathrm{[m\text{·}rad]}} & =\left\langle \mathscr{H}_{x}\left(s\right)d_{J_{t}}\left(s\right)\right\rangle =\frac{2T_{0}\varepsilon_{x}}{\tau_{x}}=\frac{C_{\mathrm{q}}C_{\gamma}\gamma^{2}E_{0}^{3}I_{5}}{2\pi}\\
D_{J_{y}}\:\mathrm{\mathrm{[m\text{·}rad]}} & =0\\
D_{J_{t}}\:\mathrm{\mathrm{[m\text{·}rad]}} & =\left\langle d_{J_{t}}\left(s\right)\right\rangle =\frac{2T_{0}\varepsilon_{s}}{\tau_{t}}=\frac{2T_{0}\sigma_{p_{t}p_{t}}}{\gamma_{t}\tau_{p_{t}}}=\frac{C_{\mathrm{q}}C_{\gamma}\gamma^{2}E_{0}^{3}I_{3}}{2\pi\gamma_{t}}
\end{aligned}
\right..
\end{equation}

\subsection{Effective Emittance}

For a dispersive straight the effective emittance is
\begin{equation}
\varepsilon_{\mathrm{eff},x}=\sqrt{\sigma_{x}^{2}\sigma_{p_{x}}^{2}-\sigma_{xp_{x}}^{2}}=\varepsilon_{0,x}\sqrt{1+\frac{\mathscr{H}_{x}\sigma_{p_{t}}^{2}}{\varepsilon_{0,x}}}
\end{equation}
since the momentum spread $\sigma_{p_{t}}$ contributes to the beam
size as well
\begin{equation}
\sigma_{x}^{2}=\beta_{x}\varepsilon_{x}+\eta_{x}^{2}\sigma_{p_{t}}^{2},\qquad\sigma_{p_{x}}^{2}=\gamma_{x}\varepsilon_{x}+\left(\eta'_{x}\right)^{2}\sigma_{p_{t}}^{2},\qquad\sigma_{xp_{x}}=-\alpha_{x}\varepsilon_{x}+\eta_{x}\eta'_{x}\sigma_{p_{t}}^{2}.
\end{equation}

\subsection{Impact of Insertion Devices}

An insertion device in a dispersive straight increases the horizontal
emittance by
\begin{equation}
\frac{d\varepsilon_{x}}{\varepsilon_{k}}=\frac{d\tau_{x}}{\tau_{x}}+\frac{dD_{J_{k}}}{D_{J_{k}}}=-\frac{dj_{k}}{j_{k}}-\frac{dU_{0}}{U_{0}}+\frac{dI_{5}}{I_{5}}
\end{equation}
since
\begin{equation}
\varepsilon_{k}=\frac{\tau_{k}D_{J_{k}}}{2T_{0}},\qquad D_{J_{x}}=\frac{C_{\mathrm{q}}C_{\gamma}\gamma^{2}E_{0}^{3}I_{5}}{2\pi},\qquad\tau_{x}=\frac{2T_{0}E_{0}}{j_{k}U_{0}}
\end{equation}
and $dI_{5}$ is given by
\begin{equation}
\Delta I_{5}=\left(\frac{B_{\mathrm{w}}}{\left(B\rho\right)}\right)^{3}\frac{4L_{\mathrm{w}}\Delta\mathscr{H}_{0,x}}{3\pi}.\label{eq:ID_impact}
\end{equation}

\section{Symbols and Notations}
\begin{elabeling}{00.00.0000}
\item [{Roman:}]~
\begin{elabeling}{00.00.0000}
\item [{$a_{n},b_{n}$}] multipole expansion
\item [{$E,B$}] electric \& magnetic fields
\item [{$C$}] circumference
\item [{$E_{0}$}] beam energy
\item [{$f_{\mathrm{RF}}$}] RF frequency
\item [{$h$}] harmonic number
\item [{$J$}] action coordinate
\item [{$L$}] element length
\item [{$p$}] particle momentum
\item [{$\left[x,p_{x},y,p_{y},c_{0}t,-p_{t}\right]$}] phase space coordinates
\item [{$T_{0}$}] particle revolution time
\item [{$v$}] particle velocity
\item [{$V_{0}$}] cavity voltage
\end{elabeling}
\item [{Greek:}]~
\begin{elabeling}{00.00.0000}
\item [{$\alpha,\beta,\gamma$}] relativistic kinimatic factors
\item [{$\varphi$}] angle coordinate
\item [{$\phi$}] relative phase deviation to $\phi_{0}$,\\
dipole bend angle, betatron motion phase
\item [{$\phi_{0}$}] RF synchronous phase
\item [{$\xi_{x,y}$}] horizontal/vertical chromaticity
\item [{$\nu_{x,y,z}$}] horizontal/vertical/longitudinal synchrotron tunes
\item [{$\omega_{0}$}] angular revolution frequency
\item [{$\tau_{x,y,z}$}] horizontal/vertical/longitudinal damping times
\end{elabeling}
\end{elabeling}

\end{document}